\documentclass[aps,prb,twocolumn,superscriptaddress,10pt]{revtex4-2}
\usepackage[utf8]{inputenc}
\usepackage{libertine}
\usepackage[libertine]{newtxmath}
\usepackage{mathtools,graphicx,microtype,bm}
\usepackage[cal=boondoxo,frak=boondox]{mathalfa}
\usepackage[colorlinks=true,allcolors=blue]{hyperref}

\begin{document}

\title{Conductance suppression by nonmagnetic point defects in helical edge channels of two-dimensional topological insulators}

\author{Vladimir A.\ Sablikov}
\email[E-mail:]{sablikov@gmail.com}
\author{Aleksei A.\ Sukhanov}
\affiliation{Kotel'nikov Institute of Radio Engineering and Electronics,
Russian Academy of Sciences, Fryazino branch, Fryazino, Moscow District, 141190, Russia}

\begin{abstract} 
We study backscattering of electrons and conductance suppression in a helical edge channel in two-dimensional topological insulators with broken axial spin symmetry in the presence of nonmagnetic point defects that create bound states. In this system the tunneling coupling of the edge and bound states results in the formation of composite helical edge states in which all four partners of both Kramers pairs of the conventional helical edge states and bound states are mixed. The backscattering is considered as a result of inelastic two-particle scattering of electrons, which are in these composite states. Within this approach we find that sufficiently strong backscattering occurs even if the defect creates only one energy level. The effect is caused by electron transitions between the composite states with energy near the bound state level. We study the deviation from the quantized conductance due to scattering by a single defect as a function of temperature and Fermi level. The results are generalized to the case of scattering by many different defects with energy levels distributed over the band gap. In this case, the conductance deviation turns out to be quite strong and comparable with experiment even at a sufficiently low density of defects. Interestingly, under certain conditions, the temperature dependence of the conductance deviation becomes very weak over a wide temperature range.

\end{abstract}

\maketitle

\section{Introduction}

The quantized conductance in the edge channel is a distinctive feature of the helical edge states (HESs) in two-dimensional (2D) topological insulators (TIs), but in real systems, the quantization of edge conductivity is often violated, which arouses great interest in elucidating the nature of this anomaly~\cite{GUSEV2019113701}. The HESs were predicted more than ten years ago~\cite{PhysRevLett.95.226801,PhysRevLett.95.146802,PhysRevLett.96.106802}. They are a Kramers doublet of counter-propagating, spin-polarized states with gapless spectrum~\cite{PhysRevB.73.045322,RevModPhys.82.3045,RevModPhys.83.1057,Ren_2016,bernevig2013topological}, and elastic scattering of electrons between Kramers partners is impossible due to time reversal symmetry. Soon after the discovery of the HESs~\cite{Konig766}, it turned out that the conductance was not strictly quantized, in contrast to theoretical expectations and hence the HESs are not fully protected against the backscattering even in the absence of magnetic interactions~\cite{PhysRevLett.123.047701,PhysRevX.3.021003,GUSEV2019113701}.

Over the last decade, many attempts have been undertaken to solve this problem on the basis of different models and using different approaches. Without pretending to be complete, we can single out several of the most significant areas of such research. 
It is clear that the breakdown of the conductance quantization can be caused by magnetic defects~\cite{PhysRevLett.102.256803,PhysRevLett.106.236402,PhysRevLett.111.086401,PhysRevB.93.241301}, but it is unlikely that such defects are present in typical TIs used in experiments. Keeping in mind such typical TIs, the ideas of the formation of states with spontaneously broken spin symmetry due to electron-electron (e-e) interaction are of greater interest. Spontaneous spin symmetry breaking can occur in the edge channel of 2D systems formed by a smooth confining edge potential~\cite{PhysRevLett.118.046801}, and an even more interesting situation arises in the presence of short-range non-magnetic defects located near the edges with Hubbard-like e-e interactions~\cite{PhysRevLett.122.016601}. In this case, local magnetic moments are formed near the defects, leading to backscattering of edge electrons. The conductance of such systems as a function of Fermi energy and temperature is still little studied, so it is difficult to understand whether these ideas can explain the existing experiments, but it is clear that in this case, the backscattering does not disappear even at extremely low temperatures.

An important mechanism for breaking topological protection is breaking the time reversal symmetry in inelastic processes. However, inelastic processes induced by phonons do not lead to any significant backscattering even in the presence of spin-orbit interaction (SOI), which breaks the axial spin symmetry~\cite{PhysRevLett.108.086602}. As a possible backscattering mechanism, inelastic e-e scattering is of much greater importance, since electron spins can change in this process. Violation of the conductance quantization arises due to two-particle interactions in the presence of disorder and impurities, even if the spin projection is conserved~\cite{PhysRevB.73.045322,PhysRevLett.96.106401,PhysRevB.85.235304}, but in this case the effect is rather small and strongly depends on temperature. However, in the absence of axial spin symmetry, the backscattering resulting from weak e-e interactions and an impurity potential increases significantly and has a not so strong dependence on temperature~\cite{PhysRevLett.108.156402}. The deviation from the quantized conductance due to this backscattering mechanism increases with temperature as $T^4$ even if the e-e interaction is weak. When the e-e interaction is strong enough, which is not the case in many experiments because of the large dielectric constant, the effects of the Luttinger liquid become significant. They lead to a weakening of the temperature dependence of the conductivity suppression due to interaction-induced inelastic scattering both on a separate defect~\cite{PhysRevB.86.121106} and on disorder~\cite{PhysRevB.90.075118}.

Further development of research in this direction was carried out for heavily doped materials, in which electron puddles are formed in the presence of a gate or a compensating charge of impurities~\cite{PhysRevLett.110.216402,PhysRevB.90.115309}. A feature of this system is the presence of electronic states with a discrete spectrum localized in puddles. Backscattering of electrons in edge states occurs as a result of a multistep process that includes the tunneling of an electron into a puddle, an electronic transition between discrete energy levels caused by the interaction of two electrons in the puddle, and tunneling back into the edge channel. The key role is played by the electron transition in the puddle, which, in the absence of axial spin symmetry, leads to a change in the spin state (more precisely, the Kramers index) of electrons in edge states.

Experimental studies carried out on the basis of quantum wells HgTe~\cite{PhysRevLett.123.047701,GUSEV2019113701,PhysRevB.89.125305} and heterostructures InAs/GaSb~\cite{PhysRevLett.114.096802,PhysRevLett.115.136804} clearly state the presence of two problems that the existing theories do not solve even at a qualitative level. First, this is a large value of the conductivity suppression effect. The measured conductance can be several or more times less than the quantum $e^2/h$ at a source-drain distance of several microns. As far as we know, there are still no convincing estimates that could quantitatively explain the magnitude of the effect for specific materials based on their parameters. Second, the suppressed conductance is surprisingly weakly dependent on temperature down to fairly low temperatures, on the order of 20~mK for heterostructures HgTe/CdHgTe. 

This signifies that the problem requires a deeper investigation. In this regard, it should be noted that the system under study is rather complicated and the calculations of the proposed mechanisms are often based on rather crude models using phenomenological parameters that are still poorly studied for specific structures used in experiments. Of crucial importance are the parameters that determine the spin flip processes. These include, first of all, the matrix elements of the e-e interaction and especially the anisotropic components of the exchange interaction tensor, tunneling matrix elements coupling edge states and bound states at defects, charging energy, etc. In fact, for calculating the conductivity, not only the magnitude of these parameters is important, but also their dependence on the energy or the wave vector of electrons. In this paper, we will show that a correct calculation of these parameters not only significantly changes the backscattering probability, but also opens up a new possibilities for the backscattering to occur and new mechanism for the temperature dependence of conductance.

First, we study the problem of electron backscattering in HESs in a 2D TI with broken axial spin symmetry in the presence of one defect with one energy level of bound states. Luttinger liquid effects are supposed to be negligible. According to existing concepts, a defect with one energy level does not create backscattering due to bound states, since no electronic transitions between energy levels of bound states can occur, as in the case of electron puddles~\cite{PhysRevB.90.115309}. The inelastic backscattering due to the joint effect of weak e-e interactions and potential scattering by the defect is rather weak and strongly dependent on temperature~\cite{PhysRevLett.108.156402}.

We find that in fact, in this case, there is a fairly strong backscattering caused by the presence of bound states. The probability of this scattering is comparable to that of a puddle. The effect is due to the formation of composite HESs, which are formed as a result of the tunneling coupling of conventional HESs and states bound at the defect~\cite{PhysRevB.102.075434}. Backscattering occurs as a result of two-particle scattering of electrons that are just in these states. The key role in this backscattering mechanism belongs to the tunneling matrix, which describes the transitions of electrons between different Kramers partners of HESs and bound states. We calculate the backscattering probability and study the deviation from the quantized conductance as a function of temperature and Fermi level.

The results obtained for one defect with one energy level are generalized to the case when there is a set of different defects with energy levels distributed over the band gap. We show that in such a situation a sufficiently strong suppression of conductance can be achieved even when a defect density is rather low. In addition, the temperature dependence of conductance can be weak over a wide temperature range. 

The backscattering theory is constructed without any model assumptions regarding the spinor structure of wave functions, the tunneling matrix elements coupling edge and bound states, and the matrix elements of e-e interaction. Single-particle wave functions are constructed in the form of four-rank spinors without often used approximations by two-rank spinors and one-dimensional functions. In final numerical calculations, we use the Bernevig-Huges-Zhang model~\cite{Bernevig1757} with SOI caused by breaking of the inversion symmetry of the material. 

The structure of the paper is as follows. In Sec.~\ref{Sec2} we briefly describe composite HESs. Section~\ref{Sec3} presents a theory of e-e scattering developed on the basis of two-particle composite HESs. Here we also study the conductance suppression in the cases of one defect and many defects with energy levels distributed over the band gap. The main results are discussed and summarized in Sec.~\ref{Sec4}. Details of the model used in numerical calculations of the tunneling matrix and the conductance are given in the Appendix~\ref{App1}.

\section{Helical edge states coupled to a defect}\label{Sec2}

Although the HESs in TIs are robust against scattering by the potential of defects, this does not mean that they remain unchanged due to interaction with the defects. The most dramatic changes occur when HESs interact with the bound states created by these defects. Such bound states almost always arise, unless the potential of a defect is too smooth or weak. In this case, conventional HESs are coupled with bound states, which leads to the formation of composite HESs~\cite{PhysRevB.91.075412}. It is in these states that electrons are in the edge channel in the presence of defects. We will consider the backscattering process as a result of two-particle scattering of electrons, taking into account the fact that the colliding electrons are in these composite HESs. The theory of composite HESs in the case of broken axial spin symmetry was developed in Ref.~\cite{PhysRevB.102.075434}. Here we briefly outline the main results that will be needed further for calculations of the backscattering rate. 

The wave function of composite HESs $\Psi$ is constructed on the basis of the conventional HESs $|k,\sigma\rangle$ and bound states $|\lambda\rangle$
\begin{equation}
\Psi=\sum\limits_{k',\sigma'}A_{k',\sigma'}|k',\sigma'\rangle + \sum\limits_{\lambda'}B_{\lambda'}|\lambda'\rangle\,,
\end{equation}
Here $|k,\sigma\rangle$ is a four-rank spinor describing HESs with broken axial spin symmetry, $k$ is the wave vector, $\sigma=\pm$ is the Kramers degeneracy index defining right- and left-moving states. $|k,\sigma\rangle$ is a 2D wave function propagating along the edge ($x$ direction) and decaying into the bulk of the 2D TI ($y$ direction). $|\lambda\rangle$ is a four-rank spinor of bound states at a point defect located at distance $d$ from the edge. In general, bound states are characterized by the quantum number $n$ of the energy level and the Kramers index $\lambda=\pm$ defining clockwise and anticlockwise circulating states. For simplicity, we are considering only one energy level here, so $n$ takes only one value. In this way, we will focus on the effects arising from the formation of composite HESs, and will not consider electron transitions between the energy levels of the defect.

The coefficients $A_{k,\sigma}$ and $B_{\lambda}$ are determined using the methods developed for the Fano-Anderson model~\cite{PhysRev.124.1866,mahan2013many}. The wave function of a composite HES is largely determined by the tunneling matrix, which couples the edge and bound states, $w_{k,\sigma;\lambda}=\langle k,\sigma|H_T|\lambda\rangle$, where $H_T$ is the tunneling Hamiltonian.

There is a Kramers doublet of composite HESs propagating to the right and to the left with the energy $E$. At infinity $|x|\to\infty$, the composite HESs coincide up to a phase with conventional HESs. Although the momentum in composite HESs is not defined, it is convenient to introduce the quantity $\mathcal{K}$, which is uniquely related to the energy $E$ by the dispersion equation for conventional HESs. Therefore, $K$ has the meaning of the wave vector at infinity. We will assume that the energy depends approximately linearly on this momentum: $E=\hbar v\mathcal{K}$, with $v$ being the velocity.

The wave functions of the composite HESs $\Psi_{\mathcal{K},R/L}$ contain three components:
\begin{equation}
\Psi_{\mathcal{K},R/L}=\Phi_{\mathcal{K},R/L}+\Psi_{\mathcal{K},R/L}^{(prop)}+\Omega_{\mathcal{K},R/L}\,.
\label{combined_WFs}
\end{equation}

The first component is formed by the bound states and is localized in nearest vicinity of the defect, 
\begin{align}
\Phi_{\mathcal{K},R}&=B_{\mathcal{K}}(\widetilde{w}_{\mathcal{K},+;+}^*\Phi_{+}+\widetilde{w}_{\mathcal{K},+;-}^*\Phi_{-})/\sqrt{L}\label{bound_comp_wfR}\,,\\
\Phi_{\mathcal{K},L}&=B_{\mathcal{K}}(-\widetilde{w}_{\mathcal{K},+;-}^*\Phi_{-}+\widetilde{w}_{\mathcal{K},+;+}\Phi_{-})/\sqrt{L}\label{bound_comp_wfL}\,.
\end{align}
It is important that the bound-state component of each Kramers partner of the composite HESs contains both Kramers partners of the bound states with the weights that are determined by the tunneling matrix components $w_{\mathcal{K},+;+}$ and $w_{\mathcal{K},+;-}$. Here and in what follows it is convenient to use the tunneling matrix elements $\widetilde{w}_{k,\sigma;\lambda}$ renormalized so that they do not depend on the normalization length $L$ of the wave function of conventional HESs along $x$-axis,
\begin{equation}
\widetilde{w}_{k,\sigma;\lambda}=w_{k,\sigma;\lambda} \sqrt{L}\,.
\end{equation}

An important feature of the bound-state component is that it has a sharp maximum as a function of the energy near the bound state energy. This resonance is described by the factor $B_{\mathcal{K}}$,
\begin{equation}
B_{\mathcal{K}}=\frac{1}{\sqrt{\left(E-\varepsilon_0-\Sigma_{\mathcal{K}}\right)^2+\gamma_{\mathcal{K}}^2}}\,,
\end{equation}
where $\varepsilon_0$ is the energy of the bound state at the defect, $\Sigma_{\mathcal{K}}$ is the self-energy function, and $\gamma_{\mathcal{K}}$ is the resonance width.  $\Sigma_{\mathcal{K}}$ and $\gamma_{\mathcal{K}}$ read as
\begin{align}
\Sigma_{\mathcal{K}}&=\frac{1}{2\pi \hbar v}\mathcal{P}\int\limits_{-K_c}^{K_c}dk'\frac{|\widetilde{w}_{k',+;+}|^2+|\widetilde{w}_{k',+;-}|^2}{\mathcal{K}-k'}\,,\\
\gamma_{\mathcal{K}}&=\frac{|\widetilde{w}_{\mathcal{K},+;+}|^2+|\widetilde{w}_{\mathcal{K},+;-}|^2}{2 \hbar v}\,.
\end{align}
Here $\mathcal{P}$ denotes the principal value of the integral, $K_c$ is the wave vector determined by the upper limit of the energy above which HESs disappear. This limiting energy is slightly higher than the band gap edge $E_g/2$, so $K_c\approx E_g/2\hbar v$. 

The resonance energy (more precisely, the wave vector $\mathcal{K}_0$) is defined by the equation
\begin{equation}
\hbar v \mathcal{K}_0-\varepsilon_0-\Sigma_{\mathcal{K}_0}=0\,.
\end{equation}
The wave vector $\mathcal{K}_0$, which roughly defines the resonance energy, and the resonance width $\gamma_{\mathcal{K}_0}$ are important quantities that, as will be seen, largely determine the temperature dependence of the conductance suppression.

Two other components of the wave function of composite HESs $\Psi_{\mathcal{K},R/L}$ in Eq.~(\ref{combined_WFs}) are formed with the participation of the conventional HESs.

The second term in Eq.~(\ref{combined_WFs}) is a propagating component. This is the only component that does not decay at infinity. At infinity, $\Psi_{\mathcal{K},R/L}^{(prop)}$ coincides up to phase with the left- or right-moving conventional HES,
\begin{align}
\Psi_{\mathcal{K},R}^{(prop)}&\simeq e^{-i \chi_{\mathcal{K}}\,\mathrm{sgn}(x)}\Psi_{\mathcal{K},+}\,,\label{prop_comp_wfR}\\
\Psi_{\mathcal{K},L}^{(prop)}&\simeq e^{i \chi_{\mathcal{K}}\,\mathrm{sgn}(x)}\Psi_{-\mathcal{K},-}\,,\label{prop_comp_wfL} 
\end{align}
where $\Psi_{\sigma\mathcal{K},\sigma}$ is the four-rank spinor of conventional HESs.

The propagating component describes a conventional HES that falls on the defect from the left or right, acquires an additional phase $\chi_{\mathcal{K}}$ when interacting with the defect, and finally runs away from the defect. The phase is defined as follows 
\begin{equation}
\tan\chi_{\mathcal{K}}= \frac{\gamma_{\mathcal{K}}}{E-\varepsilon_0 - \Sigma_{\mathcal{K}}}\,.
\end{equation}

The appearance of the phase shift $\chi_{\mathcal{K}}$ is explained by the fact that the particle transits from the edge channel into the defect, spends some time rotating around it, and then returns back to the edge channel. 

The third component in Eq.~(\ref{combined_WFs}) is a cloud formed by the conventional HESs around the defect. The cloud arises as a result of two transitions between the edge and bound states. First, an electron goes from the right- or left-moving HES with momentum $\mathcal{K}$ into one of the two Kramers partners of the bound states. Then the electron goes back to one of the Kramers partners of the HESs. It is important that the momentum is not conserved in these transitions. Therefore, an electron can pass into many conventional states with different wave vectors. A superposition of these states forms the cloud. The general form of the cloud component of the wave function is rather complicated because of the presence of integrals over the wave vectors of the HESs that make up the cloud. But it can be simplified when the energy of the composite HES is far from the gap edges, $|\mathcal{K}|<K_c$. In this case the cloud component of the wave function is approximated as
\begin{align}
\Omega_{\mathcal{K},R}&\approx B_{\mathcal{K}}\frac{\widetilde{b}_{\mathcal{K},R}}{\pi \hbar v\sqrt{L}}\frac{\sin K_c x}{x}\,,\label{cloud_wfR}\\
\Omega_{\mathcal{K},L}&\approx B_{\mathcal{K}}\frac{\widetilde{b}_{\mathcal{K},L}}{\pi \hbar v\sqrt{L}}\frac{\sin K_c x}{x}\,,\label{cloud_wfL}
\end{align}
where $\widetilde{b}_{\mathcal{K},R(L)}$ is a function of the coordinate $y$, describing the decay of the cloud into the bulk,
\begin{widetext}
\begin{equation}
\widetilde{b}_{\mathcal{K},R}\approx \frac{1}{2K_c} \int\limits_{-K_c}^{K_c}dq\left[\frac{\widetilde{w}_{\mathcal{K},+;+}^*\left(\widetilde{w}_{q,+;+}-\widetilde{w}_{\mathcal{K},+;+}\right)+\widetilde{w}_{\mathcal{K},+;-}^*\left(\widetilde{w}_{q,+;-}-\widetilde{w}_{\mathcal{K},+;-}\right)}{\mathcal{K}-q} \widetilde{\Psi}_{q,+}
+\frac{-\widetilde{w}_{\mathcal{K},+;+}^*\widetilde{w}_{-q,+;-}^*+\widetilde{w}_{\mathcal{K},+;-}^*\widetilde{w}_{-q,+;+}^*}{\mathcal{K}+q} \widetilde{\Psi}_{q,-}\right]\,,
\end{equation}

\begin{equation}
\widetilde{b}_{\mathcal{K},L}\approx \frac{1}{2K_c} \int\limits_{-K_c}^{K_c}dq\left[\frac{\widetilde{w}_{\mathcal{K},+;+}\widetilde{w}_{q,+;-}-\widetilde{w}_{\mathcal{K},+;-}\widetilde{w}_{q,+;+}}{\mathcal{K}-q} \widetilde{\Psi}_{q,+}
+\frac{\widetilde{w}_{\mathcal{K},+;+}\left(\widetilde{w}_{-q,+;+}^*-\widetilde{w}_{\mathcal{K},+;+}^*\right)+\widetilde{w}_{\mathcal{K},+;-}\left(\widetilde{w}_{-q,+;-}^*-\widetilde{w}_{\mathcal{K},+;-}^*\right)}{\mathcal{K}+q} \widetilde{\Psi}_{q,-}\right]\,.
\end{equation}
\end{widetext}
The $y$ dependence of $\widetilde{b}_{\mathcal{K},R(L)}$ is determined by the factors $\widetilde{\Psi}_{q,\sigma}(y)$ that describe the decay of conventional HESs into the bulk~\cite{PhysRevB.102.075434},
\begin{equation}
\Psi_{k,\sigma}=\frac{1}{\sqrt{L}}\widetilde{\Psi}_{k,\sigma}(y)e^{ikx}\,.
\end{equation}

Important feature of the cloud component is that it is composed of a wide set of the conventional HESs and spatially extends over large distance from the defect.

As will be seen from what follows, for the correct calculation of the backscattering probability, not only the magnitude of the tunneling matrix elements, but also their dependence on the electron momentum is of decisive importance. Therefore, it is important to calculate correctly the tunneling matrix. The tunneling Hamiltonian was found previously~\cite{PhysRevB.102.075434}. The basic equations for calculating the tunneling matrix and an example of the dependence of matrix elements on $k$ are given in the Appendix~\ref{App1}.

\section{Electron-electron scattering in the composite helical edge states}\label{Sec3}

Inelastic scattering of electrons in the presence of a defect can be considered as two-particle scattering in a situation where both electrons are in composite HESs. Of course, if the defect creates several energy levels, inelastic scattering can also occur due to electron transitions between these levels, as in puddles, but we do not consider this possibility, assuming that the defect creates a single energy level or an energy spacing between the levels is large enough. 

In this approach, we should consider electronic transitions between two-particle states due to e-e interaction. The scattering matrix for this process is calculated using two-particle wave functions, which  allows one to correctly take into account the exchange interaction of electrons without additional model assumptions. Two-particle wave functions of non-interacting electrons with wave vectors $\mathcal{K}_1$ and $\mathcal{K}_2$ are
\begin{multline}
\Psi_{\mathcal{K}_1,\nu_1;\mathcal{K}_2,\nu_2}(1,2)=\frac{1}{\sqrt{2}}\left[\Psi_{\mathcal{K}_1,\nu_1}(1)\otimes \Psi_{\mathcal{K}_2,\nu_2}(2) \right.\\ \left.- \Psi_{\mathcal{K}_2,\nu_2}(1)\otimes \Psi_{\mathcal{K}_1,\nu_1}(2)\right]\,,
\end{multline}
where $\nu$ denotes right- and left-moving states, i.e. $\nu$ is $R$ or $L$.

Using this wave function, the matrix element of the transition $|\mathcal{K}_1,\nu_1;\mathcal{K}_2,\nu_2\rangle \to |\mathcal{K}_1',\nu_1';\mathcal{K}_2',\nu_2'\rangle$ is presented as a sum of the matrix elements of direct and exchange interaction
\begin{equation}
\begin{split}
\langle\mathcal{K}_1',\nu_1',\mathcal{K}_2',\nu_2'|U(1,2)|\mathcal{K}_1,\nu_1,\mathcal{K}_2,\nu_2\rangle \\=\mathcal{M}_{\mathcal{K}_1',\nu_1';\mathcal{K}_2',\nu_2'}^{\mathcal{K}_1,\nu_1;\mathcal{K}_2,\nu_2} - \mathcal{M}_{\mathcal{K}_1',\nu_1';\mathcal{K}_2',\nu_2'}^{\mathcal{K}_2,\nu_2;\mathcal{K}_1,\nu_1}\,,
\end{split}
\label{matrix_element}
\end{equation}
where $U(1,2)=U(\mathbf{r}_1-\mathbf{r}_2)$ is the electron-electron interaction potential. The matrix elements $\mathcal{M}$ in the right are greatly simplified in the case of a short-range interaction potential approximated as $U(1,2)=U\delta(x_1-x_2)\delta(y_1-y_2)$,
\begin{equation}
\mathcal{M}_{\mathcal{K}_1',\nu_1';\mathcal{K}_2',\nu_2'}^{\mathcal{K}_1,\nu_1;\mathcal{K}_2,\nu_2}=U\!\int \!\!dx dy \left(\Psi_{\mathcal{K}_1',\nu_1'}^+\Psi_{\mathcal{K}_1,\nu_1}\right)\left(\Psi_{\mathcal{K}_2',\nu_2'}^+\Psi_{\mathcal{K}_2,\nu_2}\right)\,,   
\label{matrix_dir-exch}
\end{equation}
where the wave functions are defined by Eq.~(\ref{combined_WFs}). 

An explicit expression for the matrix element is very cumbersome, since each one-particle wave function contains three components, but it is actually simplified if we take into account that numerous terms arising from the multiplication of the wave functions are of different order of magnitude. Indeed, the procedure for calculating one-particle wave functions assumes a weak coupling between the edge and bound states, and therefore the matrix elements $w_{k,\sigma;\lambda}$ should be considered as small quantities. Thus, the total expression for the matrix element $\mathcal{M}$ is a series of terms of different order in $w_{k,\sigma;\lambda}$. Omitting the terms of higher orders, the matrix element can be represented in the form
\begin{equation}
\mathcal{M} = \mathcal{M}_{\Phi} + \mathcal{M}_{\Psi} + \mathcal{M}_{\Psi \Omega}\,,
\label{3_component_M}
\end{equation}
where $\mathcal{M}_{\Phi}$ is formed by the bound components of the wave functions, $\mathcal{M}_{\Psi}$ is formed by propagating edge wave functions, and $\mathcal{M}_{\Psi \Omega}$ is composed of products of three propagating one-particle wave functions and one cloud wave function.  

Two types of e-e scattering processes are possible in which one or both electrons are scattered back. Here we restrict ourselves to the first case, when only one electron is scattered back, such as $\Psi_{\mathcal{K}_1,R;\mathcal{K}_2,R} \to \Psi_{\mathcal{K}_1',R;\mathcal{K}_2',L}$. Backscattering of two electrons seems less probable for the real strength of SOI, but in any case, this issue requires further study.

\subsection{Backward scattering of one of two colliding electrons}

In this section we consider a scattering process in which two electrons moving to the right with momenta $\mathcal{K}_1$ and $\mathcal{K}_2$ are scattered into a state in which one electron moves to the right and the other to the left with momenta $\mathcal{K}_1'$ and $\mathcal{K}_2'$. Calculation of all three components $\mathcal{M}_{\Phi}$, $\mathcal{M}_{\Psi}$, and $\mathcal{M}_{\Psi \Omega}$ of the matrix element of this transition leads to the following results.

The bound-state component is
\begin{multline}
\mathcal{M}_{\Phi}=L^{-2}B_{\mathcal{K}_1'}B_{\mathcal{K}_1}B_{\mathcal{K}_2'}B_{\mathcal{K}_2}\widetilde{w}_{\mathcal{K}_1',+;+}\widetilde{w}_{\mathcal{K}_2',+;+}^*\widetilde{w}_{\mathcal{K}_1,+;+}^*\widetilde{w}_{\mathcal{K}_2,+;+}^* \\ \times \left[1+\frac{\widetilde{w}_{\mathcal{K}_1',+;-}\widetilde{w}_{\mathcal{K}_2',+;-}^*}{\widetilde{w}_{\mathcal{K}_1',+;+}\widetilde{w}_{\mathcal{K}_2',+;+}^*} \right]\left[\frac{\widetilde{w}_{\mathcal{K}_2,+;-}^*}{\widetilde{w}_{\mathcal{K}_2,+;+}^*}-\frac{\widetilde{w}_{\mathcal{K}_1,+;-}^*}{\widetilde{w}_{\mathcal{K}_1,+;+}^*}\right]l_{\Phi}^{-2} \,,
\label{M_Phi}
\end{multline}
where 
\begin{equation}
l_{\Phi}^{-2}=\int\!d^2r\left(\Phi_+^+\Phi_+\right)^2
\end{equation}
is a factor which is determined by the localization length of the bound states.

The propagating-state component contains two parts
\begin{equation}
\mathcal{M}_{\Psi}=\mathcal{M}_{\Psi}^{(0)}+\mathcal{M}_{\Psi}^{(1)}\,.
\label{M_psi}
\end{equation}
The first term describes a process very similar to ordinary electron scattering, which occurs even without defects~\cite{PhysRevB.73.045322}, but here there is an essential feature caused by the presence of a defect. Calculation using Eqs.~(\ref{prop_comp_wfR}), (\ref{prop_comp_wfL}), (\ref{matrix_dir-exch}), and (\ref{matrix_element}) gives 
\begin{multline}
\mathcal{M}_{\Psi}^{(0)}= L^{-1}\mathfrak{F}(\mathcal{K}_1',\mathcal{K}_1;\mathcal{K}_2',\mathcal{K}_2) \\ \times \cos(\chi_{\mathcal{K}_1'} - \chi_{\mathcal{K}_1}-\chi_{\mathcal{K}_2'}-\chi_{\mathcal{K}_2}) \delta_{\mathcal{K}_1',\mathcal{K}_1+\mathcal{K}_2+\mathcal{K}_2'}\,,
\label{M_psi^0}
\end{multline}
where the function $\mathfrak{F}$ is determined by nonorthogonality of the Kramers partners of the HESs with different wave vectors,  
\begin{multline}
\mathfrak{F}(\mathcal{K}_1',\mathcal{K}_1;\mathcal{K}_2',\mathcal{K}_2)=\!\int\limits_0^{\infty}\!\! dy \left[\left(\widetilde{\Psi}_{\mathcal{K}_1',+}^+\widetilde{\Psi}_{\mathcal{K}_1,+}\right)\left(\widetilde{\Psi}_{-\mathcal{K}_2',-}^+\widetilde{\Psi}_{\mathcal{K}_2,+}\right)\right.\\-\left.\left(\widetilde{\Psi}_{\mathcal{K}_1',+}^+\widetilde{\Psi}_{\mathcal{K}_2,+}\right)\left(\widetilde{\Psi}_{-\mathcal{K}_2',-}^+\widetilde{\Psi}_{\mathcal{K}_1,+}\right)\right]\,.
\label{mathfrakF}
\end{multline}
Obviously, $\mathfrak{F}$ is nonzero only if the axial spin symmetry is broken. Such a factor exists even without defect. 

The presence of a defect manifests itself in the cosine factor in Eq.~(\ref{M_psi^0}), which describes how the defect affects the scattering of the propagating component of the wave function. Analysis shows that this factor sharply drops, when the energy of any of the one-particle conventional HESs before and after collision, i.e. $\hbar v \mathcal{K}_1$, $\hbar v\mathcal{K}_2$, $\hbar v\mathcal{K}_1'$, or $\hbar v\mathcal{K}_2'$, is close to the resonance.  

The Kronecker's $\delta$ symbol shows that the momentum of the pair is conserved. Since energy is also conserved, both of these conservation laws impose severe restrictions on momenta. It is easy to show that the scattering process we are considering is possible only when $\mathcal{K}_2'=0$, in other words, after the scattering, one of the electrons must end up at the Dirac point, which, therefore, should be unoccupied. For this reason, this process leads to a strong temperature dependence of the backscattering rate.

The second term in Eq.~(\ref{M_psi}) is free from this restriction,
\begin{multline}
\mathcal{M}_{\Psi}^{(1)}=L^{-1}
\frac{\mathrm{sgn}(-\mathcal{K}_1'+\mathcal{K}_1+\mathcal{K}_2+\mathcal{K}_2')}{\sqrt{1+(-\mathcal{K}_1'+\mathcal{K}_1+\mathcal{K}_2+\mathcal{K}_2')^2(L/2)^2}}\\
\times 
\mathfrak{F}(\mathcal{K}_1',\mathcal{K}_1;\mathcal{K}_2',\mathcal{K}_2)\;\sin(-\chi_{\mathcal{K}_1'}+\chi_{\mathcal{K}_1}+\chi_{\mathcal{K}_2'}+\chi_{\mathcal{K}_2})\,.
\end{multline}
It describes a process in which the momentum of a pair of interacting electrons is not conserved. Moreover, this term turns to zero when $\mathcal{K}_1+\mathcal{K}_2=\mathcal{K}_1'-\mathcal{K}_2'$, but outside this point, it is this term that describes the scattering caused by the propagating components of the wave function in the presence of a defect.

The cloud component of the matrix element $\mathcal{M}_{\Psi \Omega}$ is
\begin{widetext}
\begin{align}
 \mathcal{M}_{\Psi \Omega} \approx & (\hbar vL^2)^{-1} \left[B_{\mathcal{K}_1}\cos(\chi_{\mathcal{K}_1'})\cos(\chi_{\mathcal{K}_2'})\cos(\chi_{\mathcal{K}_2})\, \mathfrak{G}_1(\mathcal{K}_1',\mathcal{K}_1;\mathcal{K}_2',\mathcal{K}_2) + B_{\mathcal{K}_2}\cos(\chi_{\mathcal{K}_1'})\cos(\chi_{\mathcal{K}_1})\cos(\chi_{\mathcal{K}_2'})\, \mathfrak{G}_2(\mathcal{K}_1',\mathcal{K}_1;\mathcal{K}_2',\mathcal{K}_2)\right. \nonumber\\ 
 &\left.+ B_{\mathcal{K}_1'}\cos(\chi_{\mathcal{K}_1})\cos(\chi_{\mathcal{K}_2'})\cos(\chi_{\mathcal{K}_2})\, \mathfrak{G}_3(\mathcal{K}_1',\mathcal{K}_1;\mathcal{K}_2',\mathcal{K}_2) + B_{\mathcal{K}_2'}\cos(\chi_{\mathcal{K}_1'})\cos(\chi_{\mathcal{K}_1})\cos(\chi_{\mathcal{K}_2})\, \mathfrak{G}_4(\mathcal{K}_1',\mathcal{K}_1;\mathcal{K}_2',\mathcal{K}_2)\right]\,,
\label{M_PsiOmega}
\end{align}
where
\begin{align}
\mathfrak{G}_1(\mathcal{K}_1',\mathcal{K}_1;\mathcal{K}_2',\mathcal{K}_2)=&\int dy \left[
\left(\widetilde{\Psi}_{\mathcal{K}_1',+}^+\widetilde{b}_{\mathcal{K}_1,R}\right)\left(\widetilde{\Psi}_{-\mathcal{K}_2',-}^+\widetilde{\Psi}_{\mathcal{K}_2,+}\right) - 
\left(\widetilde{\Psi}_{\mathcal{K}_1',+}^+\widetilde{\Psi}_{\mathcal{K}_2,+}\right)\left(\widetilde{\Psi}_{-\mathcal{K}_2',-}^+\widetilde{b}_{\mathcal{K}_1,R}\right) \right],\label{mathfrakG1}\\
\mathfrak{G}_2(\mathcal{K}_1',\mathcal{K}_1;\mathcal{K}_2',\mathcal{K}_2)=&\int dy \left[
\left(\widetilde{\Psi}_{\mathcal{K}_1',+}^+\widetilde{\Psi}_{\mathcal{K}_1,+}\right)\left(\widetilde{\Psi}_{-\mathcal{K}_2',-}^+\widetilde{b}_{\mathcal{K}_2,R}\right) - 
\left(\widetilde{\Psi}_{\mathcal{K}_1',+}^+\widetilde{b}_{\mathcal{K}_2,R}\right)\left(\widetilde{\Psi}_{-\mathcal{K}_2',-}^+\widetilde{\Psi}_{\mathcal{K}_1,+}\right) \right],\label{mathfrakG2}\\
\mathfrak{G}_3(\mathcal{K}_1',\mathcal{K}_1;\mathcal{K}_2',\mathcal{K}_2)=&\int dy \left[
\left(\widetilde{b}_{\mathcal{K}_1',R}^+\widetilde{\Psi}_{\mathcal{K}_1,+}\right)\left(\widetilde{\Psi}_{-\mathcal{K}_2',-}^+\widetilde{\Psi}_{\mathcal{K}_2,+}\right) - 
\left(\widetilde{b}_{\mathcal{K}_1',R}^+\widetilde{\Psi}_{\mathcal{K}_2,+}\right)\left(\widetilde{\Psi}_{-\mathcal{K}_2',-}^+\widetilde{\Psi}_{\mathcal{K}_1,+}\right) \right], \label{mathfrakG3}\\
\mathfrak{G}_4(\mathcal{K}_1',\mathcal{K}_1;\mathcal{K}_2',\mathcal{K}_2)=&\int dy \left[
\left(\widetilde{\Psi}_{\mathcal{K}_1',+}^+\widetilde{\Psi}_{\mathcal{K}_1,+}\right)\left(\widetilde{b}_{\mathcal{K}_2',L}^+\widetilde{\Psi}_{\mathcal{K}_2,+}\right) - 
\left(\widetilde{\Psi}_{\mathcal{K}_1',+}^+\widetilde{\Psi}_{\mathcal{K}_2,+}\right)\left(\widetilde{b}_{\mathcal{K}_2',L}^+\widetilde{\Psi}_{\mathcal{K}_1,+}\right) \right]\,.
\label{mathfrakG4}
\end{align}
\end{widetext}
This component contains both the resonance, presented by the factors $B_{\mathcal{K}}$, and the tunneling matrix elements of the transitions between different Kramers partners of edge and bound states in a wide range of momenta.

\subsection{Backscattered current}

In this section the conductance suppression is explored using the scattering approach~\cite{imry2002introduction}. We consider electron transport between the source and drain, to which a voltage is applied. The source and drain are assumed to be reservoirs in which electrons are in equilibrium. The passage of electrons in the gap between reservoirs is described by the scattering matrix. Within the framework of this approach, the excess energy arising in inelastic scattering processes is dissipated in the reservoirs.

The probability $W$ of the backscattering process $|\mathcal{K}_1,R;\mathcal{K}_2,R\rangle \to |\mathcal{K}_1',R;\mathcal{K}_2',L\rangle$ per unit of time is calculated in the Born approximation
\begin{equation}
\frac{dW}{dt}=\frac{2\pi}{\hbar}|\mathcal{M}|^2\delta[\hbar v (\mathcal{K}_1+\mathcal{K}_2-\mathcal{K}_1'-\mathcal{K}_2')]\,.
\end{equation}

To begin, we study the case where there is only one defect. According to Eq.~(\ref{3_component_M}),
\begin{equation}
|\mathcal{M}|^2\simeq |\mathcal{M}_{\Phi}|^2+|\mathcal{M}_{\Psi}^{(0)}|^2+|\mathcal{M}_{\Psi}^{(1)}|^2+|\mathcal{M}_{\Psi \Omega}|^2+2\,\mathrm{Re}[\mathcal{M}_{\Psi}^{(0)}\mathcal{M}_{\Psi \Omega}]\,.
\label{M2_components}
\end{equation}
All components present here were defined above.

The backscattering reduces the current in the edge channel from universal value $I_0=G_0 V$, where $V$ is a source-drain voltage and $G_0=e^2/(2\pi \hbar)$. 

The backscattered current can be written as
\begin{equation}
I_{bs}=e\!\sum_{\mathcal{K}_1,\mathcal{K}_2,\mathcal{K}_1',\mathcal{K}_2'}\!\left(\frac{dW}{dt}\right) \mathcal{F}(\mathcal{K}_1,\mathcal{K}_2,\mathcal{K}_1',\mathcal{K}_2')\,,
\label{Ibs}
\end{equation}
where 
\begin{multline}
\mathcal{F}(\mathcal{K}_1,\mathcal{K}_2,\mathcal{K}_1',\mathcal{K}_2')=\\ =f_{\frac{V}{2}}(\mathcal{K}_1)f_{\frac{V}{2}}(\mathcal{K}_2)\left[1-f_{\frac{V}{2}}(\mathcal{K}_1')\right]\left[1-f_{-\frac{V}{2}}(\mathcal{K}_2')\right]\\
-f_{-\frac{V}{2}}(\mathcal{K}_1)f_{-\frac{V}{2}}(\mathcal{K}_2)\left[1-f_{-\frac{V}{2}}(\mathcal{K}_1')\right]\left[1-f_{\frac{V}{2}}(\mathcal{K}_2')\right]\,.
\label{4-distr_func}
\end{multline}
The function $\mathcal{F}$ is determined by the filling of the right- and left-moving single-particle states of the colliding electrons with Fermi levels shifted by $\pm V/2$. $f_{V}(\mathcal{K})$ is the Fermi function,
\begin{equation}
f_{V}(\mathcal{K})=\left[1+\exp{\left(\frac{\hbar v \mathcal{K}-\mu-eV}{T}\right)}\right]^{-1}\,,
\end{equation}
$\mu$ being the Fermi energy.

According to Eq.~(\ref{M2_components}), the backscattered current is presented in the form:
\begin{equation}
I_{bs}=I_{\Phi}+I_{\Psi}^{(0)}+I_{\Psi}^{(1)}+I_{\Psi \Omega}+I_{\Psi \Omega}^{(1)}\,,
\label{bs_current}
\end{equation}
where the series of current components corresponds to the components of $|\mathcal{M}|^2$ in Eq.~(\ref{M2_components}). Equation (\ref{bs_current}) together with Eqs.~(\ref{M_Phi}) -- (\ref{Ibs}) allows one to calculate the conductance deviation $\Delta G=(dI_{bs}/dV)_{V=0}$ from the quantized value $G_0$ due to backscattering by isolated nonmagnetic defects with minimal model assumptions. 

It is interesting to clarify how different components of the composite wave function~(\ref{combined_WFs}) contribute to the failure of conductance quantization. To this end, we present the complete expression for the conductance deviation $\Delta G$ as a sum of three parts in accordance with the wave function components: 
\begin{equation}
\Delta G=\Delta G_{\Phi}+\Delta G_{\Psi}+\Delta G_{\Psi\Omega}\,,
\label{DeltaG_Phi_Psi_Omega}
\end{equation}
where the bound-state component $\Delta G_{\Phi}$ stems from the current component $I_{\Phi}$, the propagating state component $\Delta G_{\Psi}$ stems from $I_{\Psi}^{(0)}+I_{\Psi}^{(1)}$, and the cloud component $\Delta G_{\Psi\Omega}$ is due to $I_{\Psi \Omega}+I_{\Psi \Omega}^{(1)}$. For convenience, in what follows we will consider the conductance deviation $\Delta G$ normalized to $G_0$. 

Below, we consider in detail all three components of the conductance deviation in Eq.~(\ref{DeltaG_Phi_Psi_Omega}). Calculations are performed numerically using the above analytical expressions. The tunneling matrix is calculated within the Bernevig-Hughes-Zhang (BHZ) model~\cite{Bernevig1757} for specific material parameters close to the HgTe/CdTe heterostructure. Details of the model and computational approaches are given in Appendix~\ref{App1}. 

Essential parameters of the model, which are used in below estimates, are: $M$ is the mass or gap parameter used to normalize the energy; $B$ is an expansion parameter of the electron and hole band energy; $a=A|B M|^{-1/2}$ is the hybridization parameter of the basis electron and hole states; $|B/M|^{1/2}$ is the length parameter; $\Delta$ is the SOI parameter. 

\subsubsection{The propagating-state component}
First consider the propagating component. The conductance deviation can be written as follows 
\begin{equation}
\frac{\Delta G_{\Psi}}{G_0}=\frac{U^2L}{4\pi \hbar v T}\int d\mathcal{K}_1 d\mathcal{K}_2 \left[\left(-\mathcal{F}'\right)|\mathfrak{F}|^2\right]_{\mathcal{K}_1,\mathcal{K}_2,\mathcal{K}_1'=\mathcal{K}_1+\mathcal{K}_2,\mathcal{K}_2'=0}\,,
\label{delta_G_psi}
\end{equation}
where $\mathcal{F}'=d\mathcal{F}/d(eV/2T)$ and the four-argument functions $\mathfrak{F}$ and $\mathcal{F}$, which are defined by Eqs.~(\ref{mathfrakF}) and (\ref{4-distr_func}), are taken with arguments $\mathcal{K}_1'=\mathcal{K}_1+\mathcal{K}_2$ and $\mathcal{K}_2'=0$.

It is seen that $\Delta G_{\Psi}$ does not depend on the defect parameters and, therefore, the presence of a defect does not affect it in any way. This is easy to understand, since the presence of a defect leads only to a phase change, which does not affect the current. The backscattering due to the e-e collisions in HESs without defects has been discussed for a long time using various approaches~\cite{PhysRevLett.95.146802,PhysRevLett.95.226801,PhysRevB.73.045322,PhysRevLett.96.106401,PhysRevB.90.075118} and is considered irrelevant to experiment because of its strong temperature dependence~\cite{PhysRevLett.108.156402}, but the value of this effect has not been estimated. Here we calculate this value and its temperature dependence. 

The backscattered current is determined mainly by electron transitions between the Kramers partners of the HESs. They are described by the function $\mathfrak{F}$, which according to Eq.~(\ref{mathfrakF}) contains products of two functions
\begin{align}
f_{k',k}(y)=&\widetilde{\Psi}_{k',+}^+\widetilde{\Psi}_{k,+},\label{f}\\
g_{k',k}(y)=&\widetilde{\Psi}_{-k',-}^+\widetilde{\Psi}_{k,+}.\label{g}
\end{align}
The function $f_{k',k}(y)$ associated with transitions between HESs with the same Kramers indexes does not have peculiarities dramatically affecting the backscattering probability. The important role belongs to the function $g_{k',k}(y)$ which reflects the possibility of transitions between the HESs with the opposite Kramers indexes, but $g_{k',k}(y)$ turns to zero at $k'=k$ due to which the integral in Eq.~(\ref{delta_G_psi}) is rather small. In particular, in the case of axial spin symmetry, the function $g_{k',k}(y)$ equals zero, whence it follows that $\Delta G_{\Psi}=0$ as well. 

But the main reason affecting the value of $\Delta G_{\Psi}$ is strong reduction of the phase volume, where the electron transitions are possible, imposed by the requirement that $\mathcal{K}_2'=0$. As a result $\Delta G_{\Psi}$ turns out to be small, but it should be noted that in the ballistic regime, this component of the conductance deviation increases with the length $L$. 

We have estimated $\Delta G_{\Psi}$ numerically for realistic conditions of the HgTe/CdTe heterostructures using two values of the hybridization parameter $a$=2 and $a$=5. The amplitude of the e-e interaction potential $U$ is estimated as follows: 
\begin{equation}
U=\int_0^{R_a}\! d^2r\frac{e^2}{\epsilon r}=\frac{2\pi e^2}{\epsilon}R_a\,,
\end{equation}
where $R_a$ is interaction radius and $\epsilon$ is the dielectric constant. If $R_a$ is taken of the order of $\sqrt{|B/M|}$ and $\epsilon\approx 20$, the potential amplitude is estimated numerically as $U\approx 10\,|B|$. 

\begin{figure}
\includegraphics[width=0.8\linewidth]{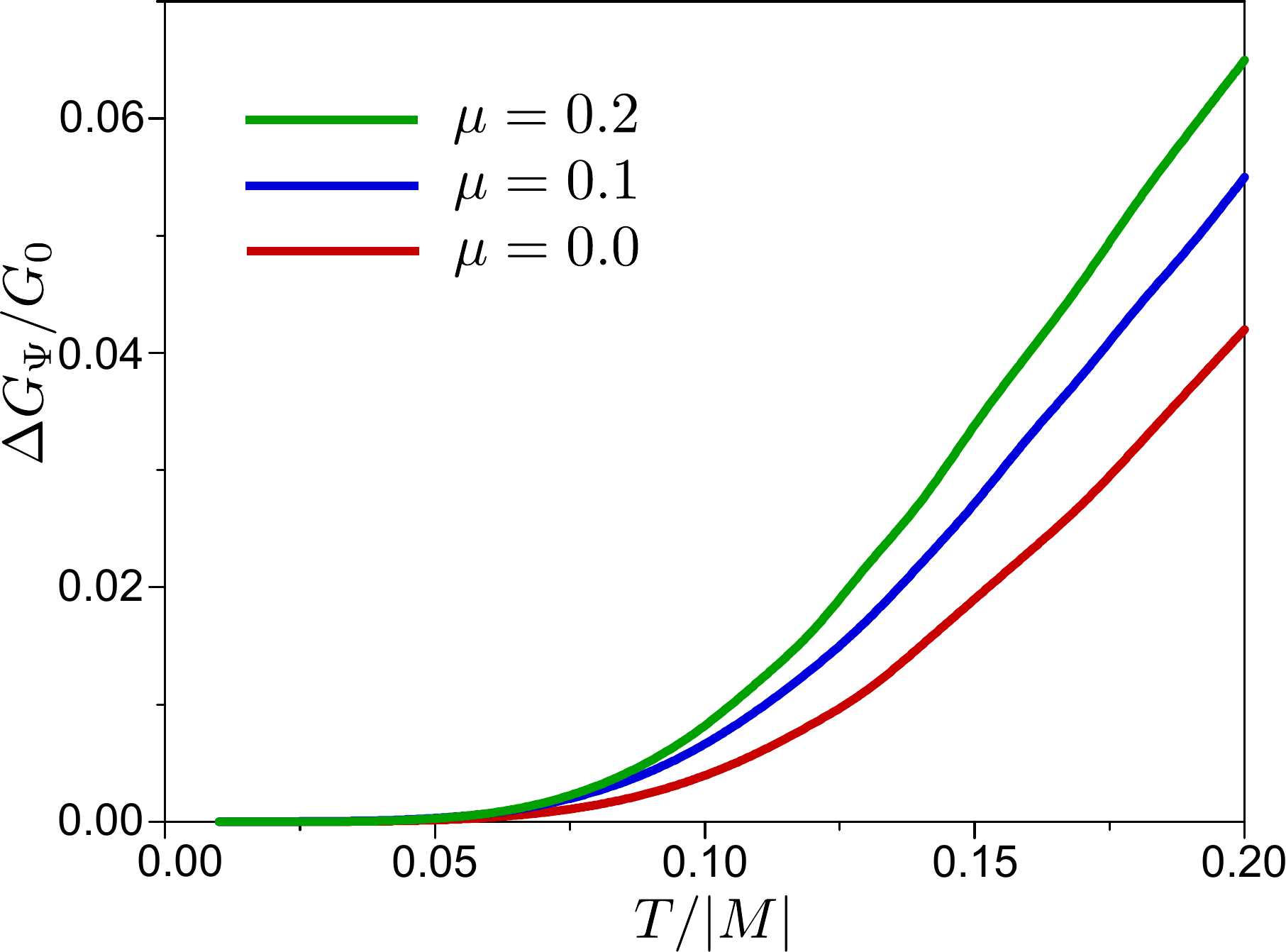}
 \caption{Conductance deviation from $G_0$ due to propagating-state component of the backscattering rate, $\Delta G_{\Psi}/G_0$, as a function of the temperature for different Fermi levels. Parameters: $a=2$, $L=10^{-2}$\,cm, $U=10\,|B|$, $\Delta=0.3\,|M|$.} 
\label{fig_1}
\end{figure}

In the case of $a=2$, the results are presented in Fig.~\ref{fig_1} for several Fermi levels. The value of $\Delta G_{\Psi}$ is seen to be small even for large source-drain distance (which is taken of the same order as in the experiments~\cite{PhysRevLett.123.047701}) and becomes essential only at very high temperature when band conductance can be important. The temperature dependence of $\Delta G_{\Psi}$ is close to $\Delta G_{\Psi}\sim T^5$ which is consistent with estimates in the framework of the Luttinger liquid model of HESs~\cite{PhysRevLett.95.226801}. In the case of a larger hybridization parameter $a=5$, which is closer to the parameters of HgTe/CdTe structures, the value of $\Delta G_{\Psi}$ decreases by approximately three orders of magnitude. Thus the propagating-state component of $\Delta G$ is small and will be ignored hereinafter.

\subsubsection{The bound-state component} 
Now turn to the bound-state component of the conductance deviation which turned out to be the most important. $\Delta G_{\Phi}$ is given by the following equation:
\begin{equation}
\frac{\Delta G_{\Phi}}{G_0}\!=\!\frac{U^2}{(2\pi)^2 \hbar v T}\!\int\! d\mathcal{K}_1 d\mathcal{K}_2 d\mathcal{K}_1' \left[\left(-\mathcal{F}'\right)|\widetilde{\mathcal{M}}_{\Phi}|^2\right],
\label{delta_G_phi}
\end{equation}
where, for convenience, we have introduced matrix element $\widetilde{\mathcal{M}}_{\Phi}$ independent of $L$: $\widetilde{\mathcal{M}}_{\Phi}=L^2\mathcal{M}_{\Phi}$, with $\mathcal{M}_{\Phi}$ being defined by Eq.~(\ref{M_Phi}). The functions of four arguments in square brackets are taken with arguments $\mathcal{K}_2'=\mathcal{K}_1 + \mathcal{K}_2- \mathcal{K}_1'$.

\begin{figure}
\includegraphics[width=0.8\linewidth]{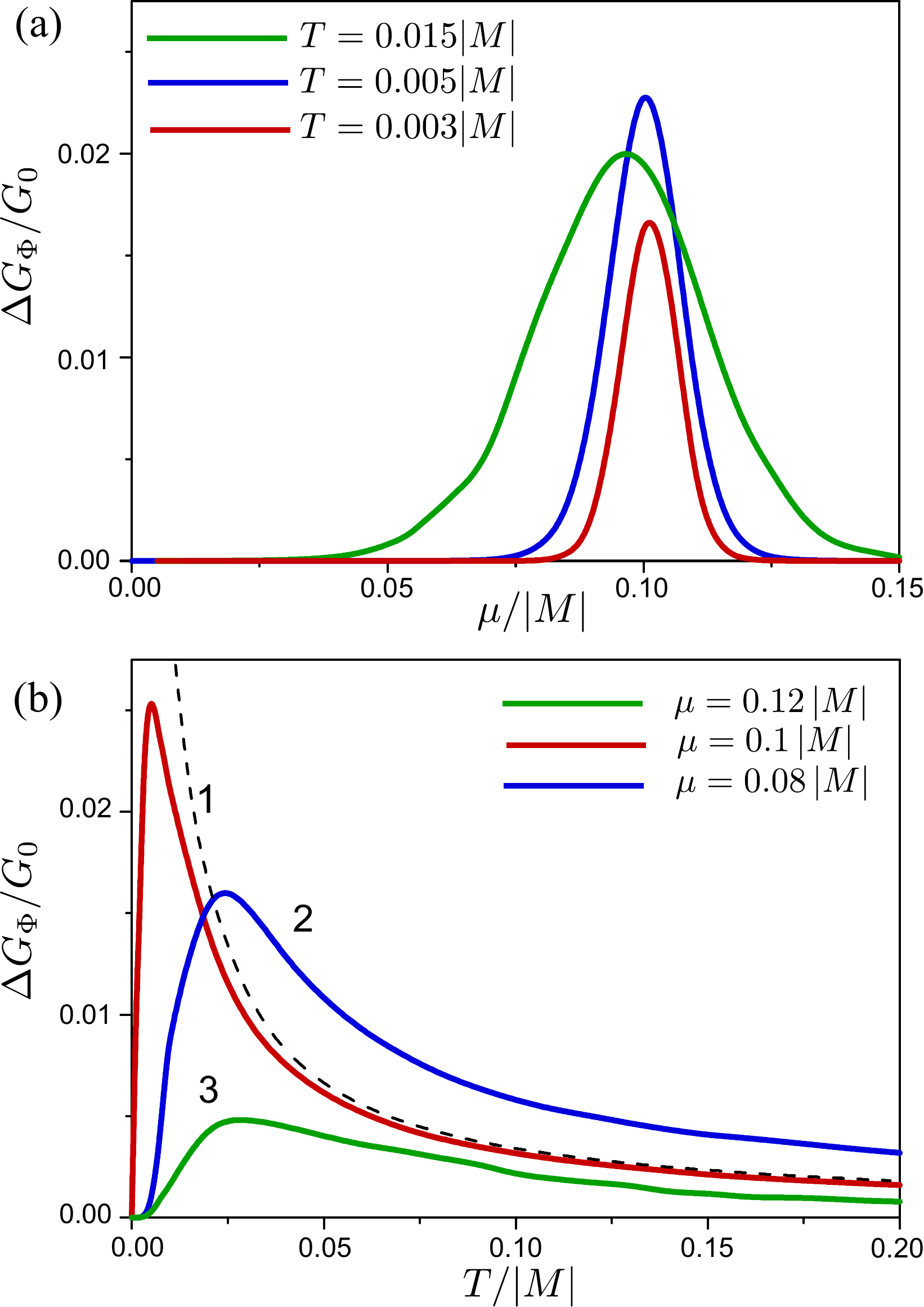}
 \caption{Conductance deviation from $G_0$ due to bound-state component of the backscattering rate, $\Delta G_{\Phi}/G_0$, as a function of (a) the Fermi energy $\mu$ for different temperatures, and (b) the temperature for different Fermi energies. The dashed line shows $1/T$ dependence. Parameters: $a=2$, $U=10\,|B|$, $\Delta=0.3\,|M|$, $\varepsilon_0=0.1\,|M|$, $d=6\,|B/M|^{1/2}$, $l_{\Phi}^{-2}=0.3\,|M/B|$.} 
\label{fig_2}
\end{figure}

In contrast to the propagating component, in this case there is no restriction imposed by momentum conservation. The integral in Eq.~(\ref{delta_G_phi}) was calculated numerically and the results were analyzed for a wide range of parameters such as the Fermi energy, the temperature, the distance $d$ between the defect and the edge, and the hybridization parameter $a$. An example of the results obtained is presented in Fig.~\ref{fig_2} where $\Delta G_{\Phi}$ is shown as a function of Fermi energy and temperature in the case when there is only one defect in the source-drain gap.

Analysis of the results of many calculations leads to the following conclusions.

First, already a single defect produces a sufficiently strong backscattering, even if it is not very close to the edge and the tunneling matrix elements are of the order of $10^{-1}$. Thus, the presence of 50 to 100 defects is sufficient to halve the conductance.

Second, the conductance deviation $\Delta G_{\Phi}$ as a function of the Fermi energy has a peak, which stems from the resonance of the composite HESs. At low temperature, $T\lesssim \gamma_{\mathcal{K}_0}$, the peak width is determined by the width of this resonance, $\gamma_{\mathcal{K}_0}$. But at larger temperature it significantly increases with temperature. 

As a function of temperature, $\Delta G_{\Phi}$ also has a maximum but of completely different form. At low temperature $T\ll \gamma_{\mathcal{K}_0}$, the conductance deviation increases as $T^4$. With increasing temperature $\Delta G_{\Phi}$ reaches a maximum, the height and width of which depend on the Fermi energy, and then decreases approximately as $T^{-1}$.

In the temperature range above $\gamma_{\mathcal{K}_0}$, which is of greatest interest in what follows, $\Delta G_{\Phi}$ is easy to estimate analytically. In this case, the function $B_{k}$ can be approximated by the delta function
\begin{equation}
B_k\simeq\frac{\pi}{\gamma_k}\delta(\hbar v k - \varepsilon_0 - \Sigma_k)\,,
\end{equation}
which shows that all four momenta involved into the scattering process are very close to the resonance value $\mathcal{K}_0$. In such a way, $\Delta G_{\Phi}$ is estimated as
\begin{equation}
\frac{\Delta G_{\Phi}}{G_0}\sim \mathcal{D}(k_0) \frac{1}{T}\frac{2e^{3(\varepsilon_0+\Sigma-\mu)/T}}{\left[1+e^{(\varepsilon_0+\Sigma-\mu)/T}\right]^{5}},
\label{delta_G_phi_simple}
\end{equation}
where
\begin{equation}
\mathcal{D}(k_0)=\frac{\pi U^2}{(\hbar v)^3 l_{\Phi}^4} \frac{|\widetilde{w}_{\mathcal{K}_0,+;+}|^4}{|\widetilde{w}_{\mathcal{K}_0,+;+}|^2+|\widetilde{w}_{\mathcal{K}_0,+;-}|^2}\left|\frac{d}{dk}\frac{\widetilde{w}_{k,+;-}}{\widetilde{w}_{k,+;+}} \right|_{k=\mathcal{K}_0}^2\,.
\end{equation}

Consequently, it follows that:\\ 
(i) $\Delta G_{\Phi}$ is a quantity of the order of $|w_{k,\sigma;\lambda}|^2$ in the series of expansion in terms of the tunneling matrix,\\
(ii) the energy dependence of the tunneling matrix is crucially important for the backscattering rate and conductance suppression,\\
(iii) when the Fermi level coincides with the resonance energy, $\Delta G_{\Phi}$ changes with temperature as $1/T$.  

\subsubsection{The cloud component}\label{SS_cloud}
The cloud component of the conductance deviation is determined by the last two terms in Eqs.~(\ref{M2_components}) and (\ref{bs_current}), however an analysis shows that the second of them is not significant since it requires conservation of the total momentum of colliding electrons, similarly to the propagating-state component. Thus the cloud component reads
\begin{equation}
\frac{\Delta G_{\Psi\Omega}}{G_0}=\frac{U^2}{4\pi^2(\hbar v)^3T}\int d\mathcal{K}_1 d\mathcal{K}_2 d\mathcal{K}_1'\left[\left(-\mathcal{F}'\right)|\widetilde{\mathcal{M}}_{\Psi\Omega}|^2\right]\,,
\end{equation}
where $\widetilde{\mathcal{M}}_{\Psi\Omega}=L^2\mathcal{M}_{\Psi\Omega}$ is independent of $L$, with $\mathcal{M}_{\Psi\Omega}$ being defined by Eq.~(\ref{M_PsiOmega}), and the functions of four arguments ($\mathcal{K}_1, \mathcal{K}_2, \mathcal{K}_1', \mathcal{K}_2'$) in brackets are taken at $\mathcal{K}_2'=\mathcal{K}_1+\mathcal{K}_2-\mathcal{K}_1'$.

\begin{figure}
\includegraphics[width=0.8\linewidth]{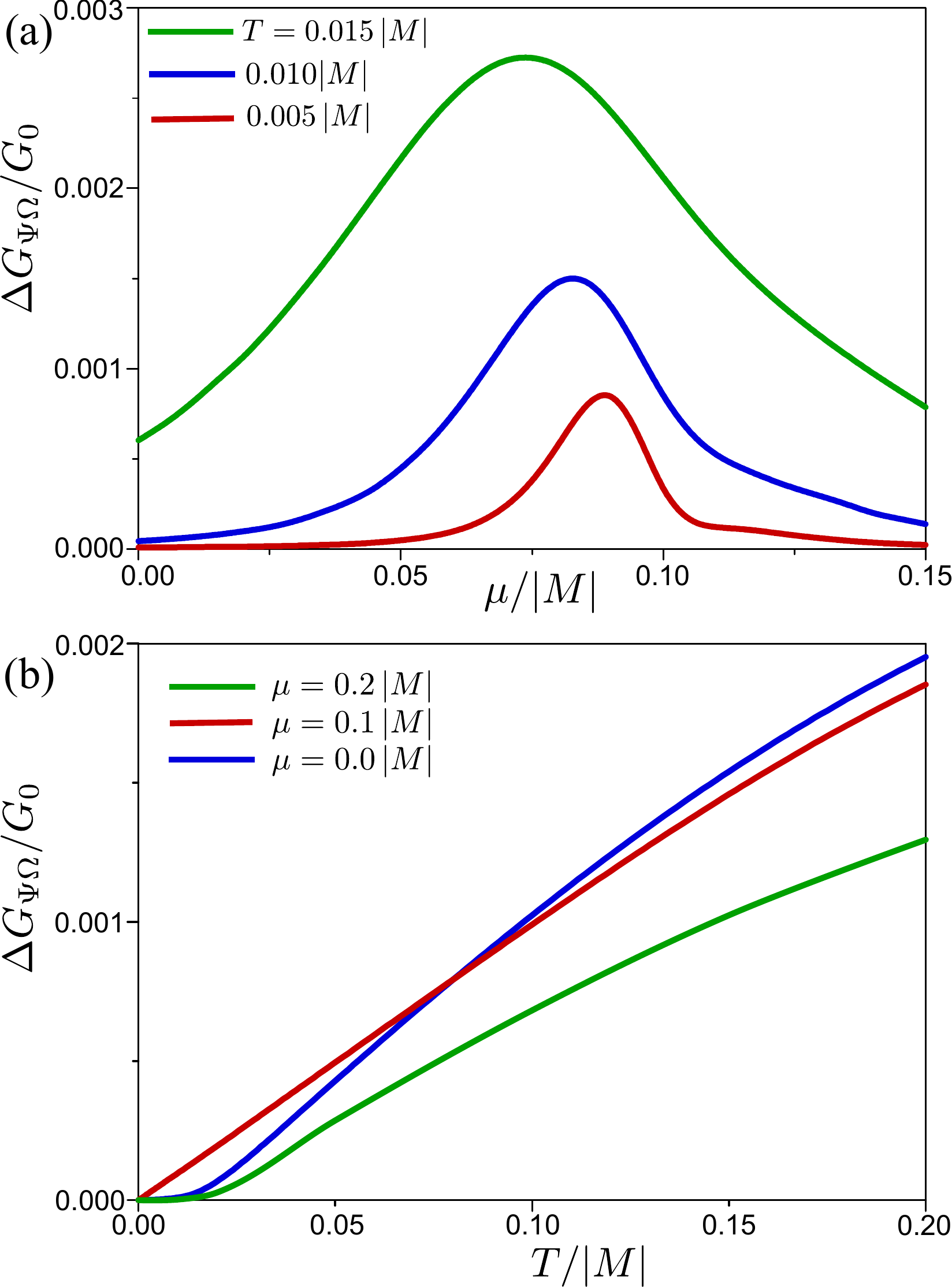}
 \caption{Conductance deviation from $G_0$ due to cloud component of the backscattering rate, $\Delta G_{\Psi\Omega}/G_0$, as a function of (a) the Fermi energy $\mu$ for different temperatures and (b) the temperature for different $\mu$. Parameters: $a=2$, $U=10\,|B|$, $\Delta=0.3\,|M|$, $\varepsilon_0=0.1\,|M|$, $d=6\,|B/M|^{1/2}$.} 
\label{fig_3}
\end{figure}

Direct calculation using Eqs.~(\ref{M_PsiOmega}) -- (\ref{mathfrakG4}) shows that $\Delta G_{\Psi\Omega}$ as a function of the Fermi energy has a maximum near the resonance of the composite HESs as shown in Fig.~\ref{fig_3} for parameters close to HgTe/CdTe heterostructures, but its width is much larger than the peak width of the bound component, for comparison see Fig.~\ref{fig_2}. An essential feature of the cloud component of the conductance deviation is that it increases significantly with temperature in contrast to the bound-state component. 

The large width of the conductance peak is obviously due to the fact that the integrand contains a resonance when integrating only over one of three momenta. For the same reason, the temperature dependence of this component of the conductance differs from the bound-state component. 

The relative contribution of the cloud and bound-state components to the total conductance deviation depends on the Fermi energy and temperature, and changes significantly with changing the model parameters of the BHZ model and the distance $d$. The calculations carried out for the material parameters close to heterostructures HgTe/CdTe lead to the following conclusions.

The cloud component becomes significant at sufficiently high temperatures.

When the Fermi energy is near the resonance, the cloud component $\Delta G_{\Psi \Omega}$ is small compared to the bound-state one $\Delta G_{\Phi}$. This is due to the fact that they are caused by different channels of electronic transitions leading a ``spin'' flip (more precisely, transitions between Kramers partners). In the case of the bound component, the ``spin'' flips as a result of two transitions between edge and bound states, which are described by the matrix $w_{k,\sigma;\lambda}$. In contrast, the ``spin'' flip with participation of the cloud occurs as a result of a combination of a transition between edge and bound states and a transition between Kramers partners of edge states, which is described by functions $g_{k',k}$. These latter transitions are much less probable than the first ones.

\begin{figure}
\includegraphics[width=0.8\linewidth]{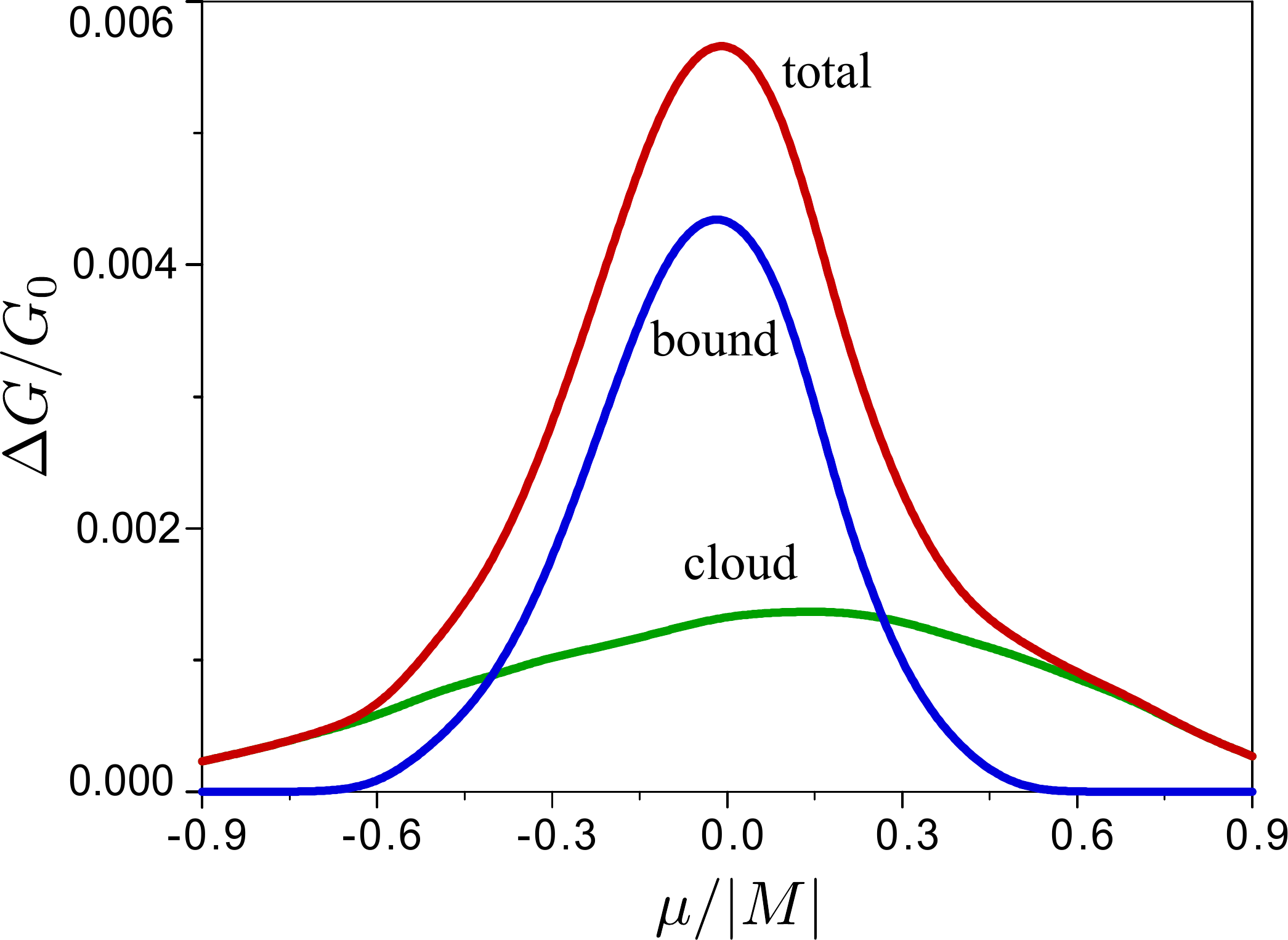}
 \caption{Total deviation of the conductance due the backscattering on the single defect and contributions of the bound-state and cloud components as functions of the Fermi energy $\mu$. Parameters: $a=2$, $U=10\,|B|$, $\Delta=0.3\,|M|$, $\varepsilon_0=0.03\,|M|$, $d=6\,|B/M|^{1/2}$, $T=0.2|M|$.}
\label{fig_4}
\end{figure}

When the Fermi level is far from the resonance the cloud component may predominate as shown in Fig.~\ref{fig_4} for the specified parameters.

The relative contribution of the cloud and bound-state components of $\Delta G$ depends also on the material parameters, among which the hybridization parameter $a$ plays an important role. A change in the parameter $a$ leads to a change in the velocity $v$, which directly affects the amplitude of the cloud component of the wave function according to Eqs.~(\ref{cloud_wfR}) and~(\ref{cloud_wfL}). As a result, an increase in $a$ leads to a decrease in $\Delta G_{\Psi\Omega}/G_0$. 

\subsection{Conductance suppression by many defects}

In reality, the materials studied in experiments, apparently contain many different defects, which can significantly change the conductance primarily due to large scatter in the defect parameters. In the model considered here, the main parameters are the energy $\varepsilon_0$ of bound states and the distance $d$ between the defect and the edge. The scatter of energy levels leads to a scatter of the resonance energies of various scatterers over the band gap, while the scatter in the positions of the defects relative to the edge leads to a scatter in the width of the resonances. In this section we explore the effect of the first factor.

We consider a 2D TI containing many different defects that are located at the same distance from the edge and create energy levels  distributed over the band gap. The density of the defects is assumed not to be very high so that they can be considered as independent scatterers. An interesting problem of correlation effects in a system of many defects requires a separate study, but something is clear already now. 

As is well known, a distinctive feature of HESs is that the scattering of electrons by the potential of the defects is strongly suppressed due to topological protection, as evidenced, for example, by the absence of Anderson localization~\cite{RevModPhys.88.035005}. So, coherent scattering effect arising from the defect potential which acts directly on the HESs is apparently small. The situation changes when the defects create bound states located at some distance from the edge, which also act on the HESs, but in this case via a tunneling coupling. This leads to a significant correlation between the various defects~\cite{PhysRevB.102.075434}. The correlation extends over large distance, and its main effect is a splitting of the resonances of the defects with close energies. The splitting energy depends on the distance between defects and can actually be on the order of a few tenths of the gap width. Therefore, one can expect that the distribution function of the resonances over their energy will be significantly smoothed due to these correlations as compared with the distribution function of the energy levels of isolated defects.

Here we restrict ourselves to a simplified approach by considering the defects as independent resonant scatterers with the energy levels $\varepsilon_0$ distributed over the band gap with the density $\rho(\varepsilon_0)$. This will allow us to estimate the magnitude of the expected backscattering effect and its temperature dependence. 

Consider a set of defects with energies $\varepsilon_0$ distributed over the band gap which are located at the same distance $d$ from the edge. The integrated effect is calculated as follows:
\begin{equation}
\Delta G=\int d\varepsilon_0\, \rho (\varepsilon_0) \Delta G(\varepsilon_0)\,,
\end{equation}
where $\Delta G(\varepsilon_0)$ is defined by Eq.~(\ref{DeltaG_Phi_Psi_Omega}) for a given $\varepsilon_0$. 

\begin{figure}
\includegraphics[width=0.8\linewidth]{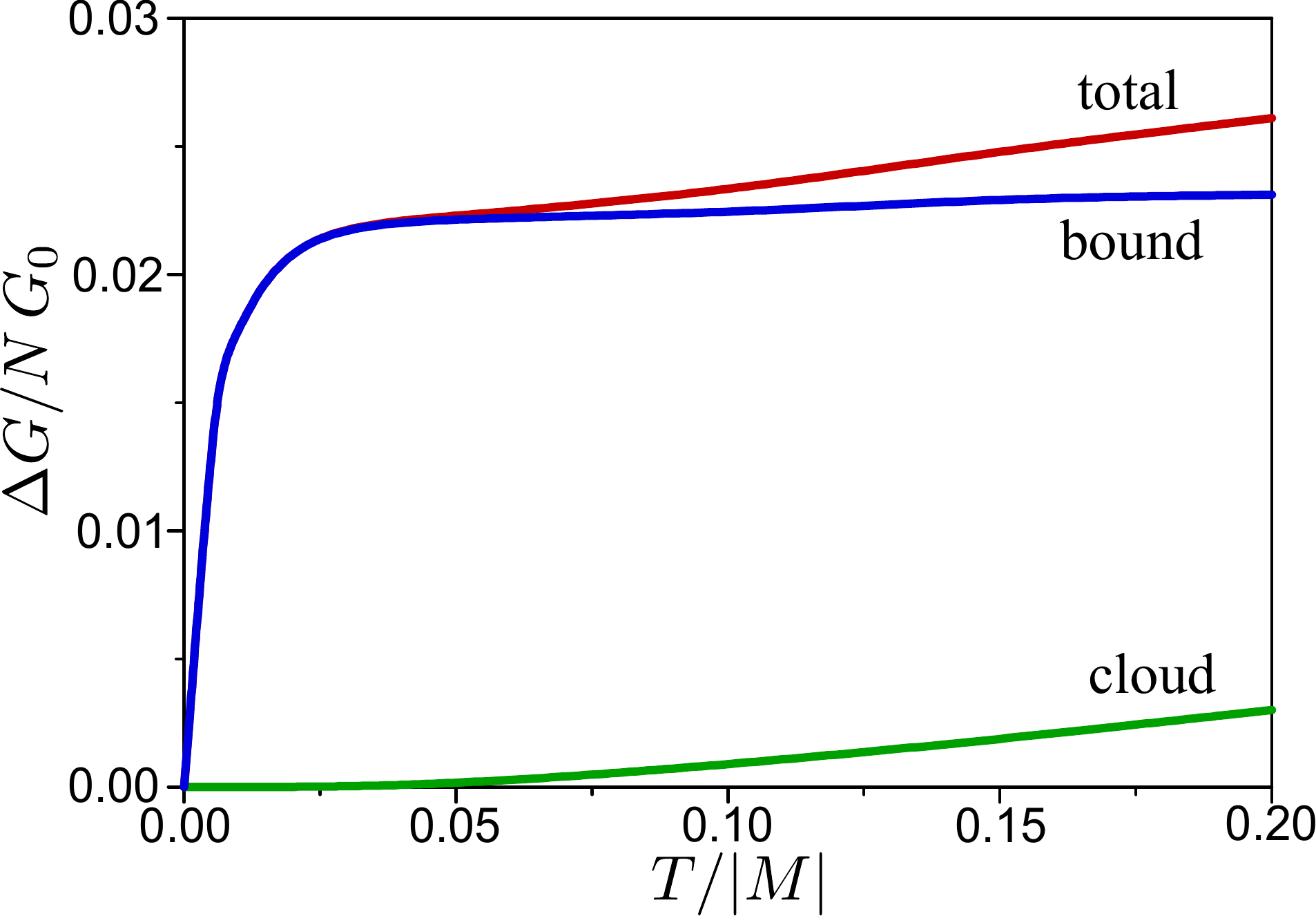}
 \caption{Total deviation of the conductance from $G_0$ due the backscattering on a set of $N$ defects with energies homogeneously distributed in the band gap as a function of temperature. Contributions of the bound-state and cloud components are shown by blue and green lines. Parameters: $a=2$, $U=10\,|B|$, $\Delta=0.3\,|M|$, $\mu_0=0.01\,|M|$, $d=6\,|B/M|^{1/2}$.}
\label{fig_5}
\end{figure}

Results of this calculation are shown in Fig.~\ref{fig_5} for homogeneous distribution of the defect energies over the band gap. It is seen that the summation over all defects radically changes the temperature dependence of the total conductance deviation. $\Delta G$ rises sharply at low temperatures, but then changes very weakly over a wide temperature range. The characteristic temperature separating these two regimes will be denoted by $T_s$.

Of greatest interest is the regime of weak temperature dependence at $T>T_s$. We have found that the temperature dependence becomes weak when the bound-state component $\Delta G_{\Phi}$ is dominant. In this case, the cloud component $\Delta G_{\Psi \Omega}$ creates a relatively small increase in the total conductance deviation with temperature. Further study has shown that the temperature dependence of $\Delta G$ changes with varying the distribution function. The function $\Delta G(T)$ can become both increasing and decreasing, but remains weakly changing if $\rho (\varepsilon_0)$ changes slowly. 

\begin{figure}
\includegraphics[width=0.8\linewidth]{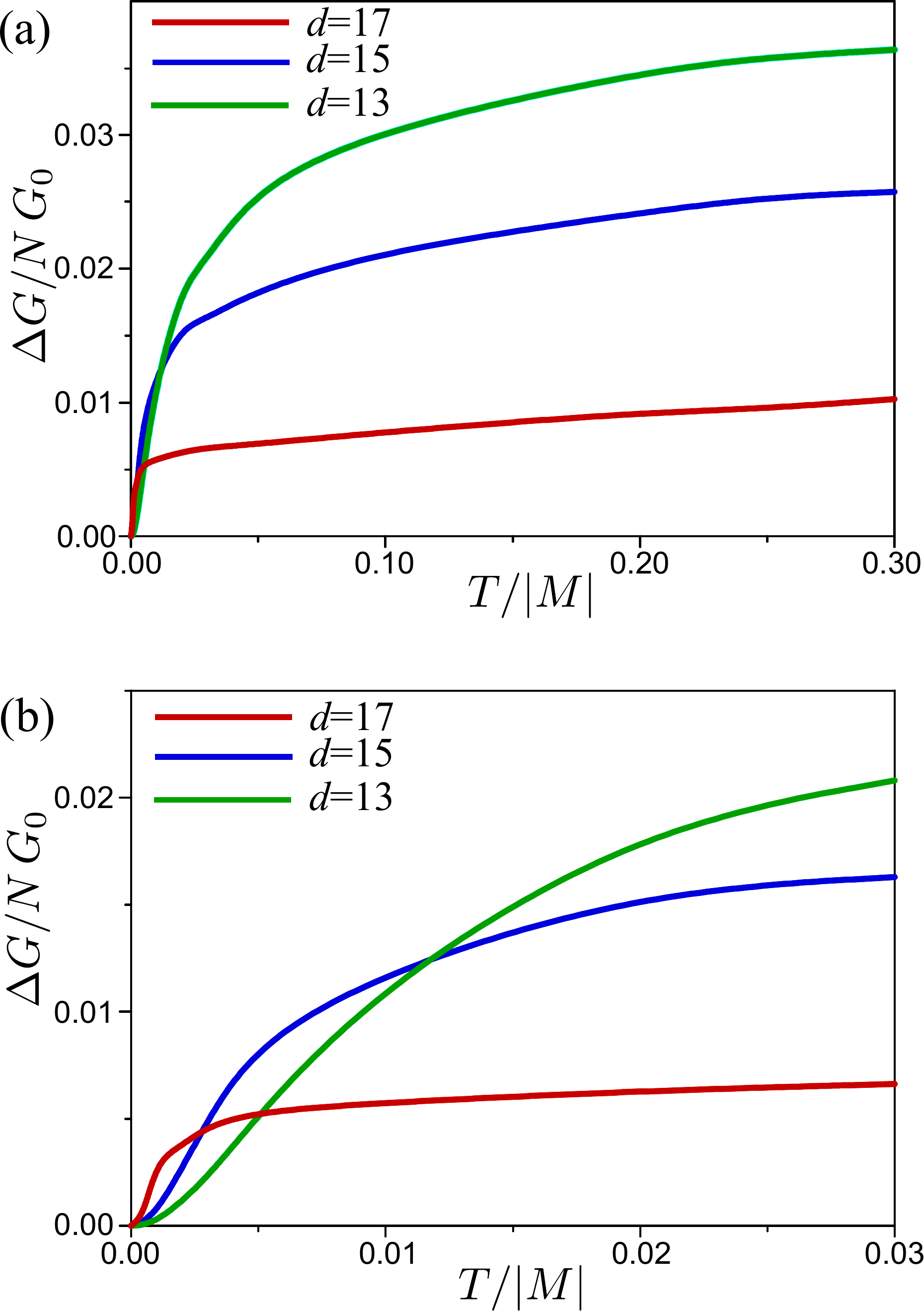}
 \caption{Total deviation of the conductance due the backscattering on a set of $N$ defects with energies homogeneously distributed in the band gap as a function of temperature for different distance $d$ between the defects and the edge, $d=13, 15, 17 \times |B/M|^{1/2}$. $\Delta G$ is normalized to the number $N$ of defects. Parameters: $a=5$, $U=10\,|B|$, $\Delta=0.3\,|M|$, $\mu=0.01\,|M|$. Panel (b) shows in more detail the low-temperature part of panel~(a).}
\label{fig_6}
\end{figure}

The physical reasons why the bound component is predominant and the conditions when the cloud component is important were discussed in the previous section. Here we study how the temperature dependence of $\Delta G$ changes with a change in the distance $d$ between defects and the edge, which strongly affects the tunneling matrix and the sharpness of resonances. The calculation has been carried out for material parameters close to HgTe/CdHgTe quantum wells with the hybridization parameter $a=5$ and homogeneous distribution of the energy levels of defects over the band gap. The results are shown in Fig.~\ref{fig_6}. It is seen that an increase in the distance $d$ leads to a decrease in the conductance deviation $\Delta G$ and the formation of a wide region of weak temperature dependence. 

\begin{figure}
\includegraphics[width=0.8\linewidth]{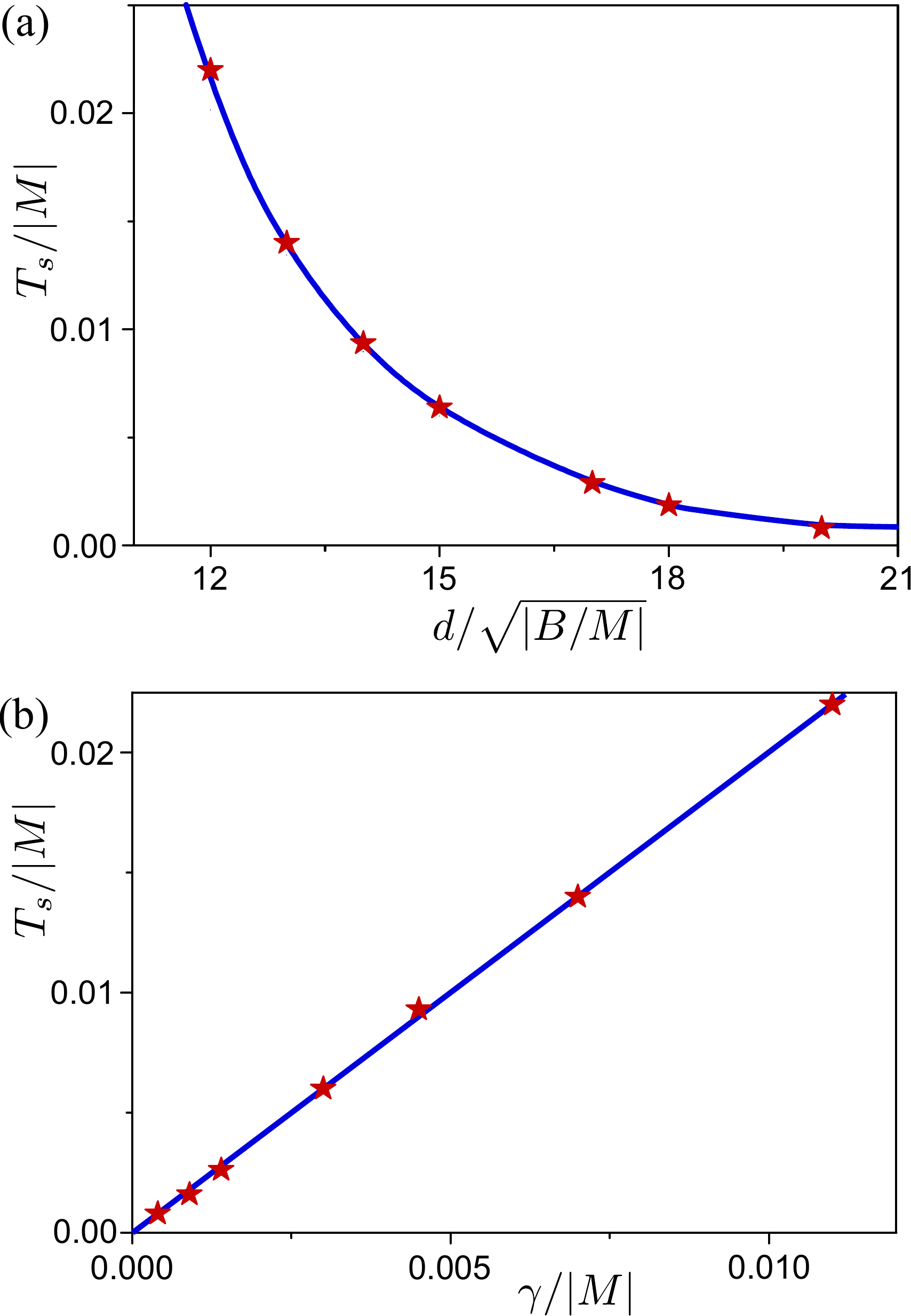}
 \caption{The characteristic temperature $T_s$ as a function of (a) the distance $d$, and (b) the width $\gamma_{k_F}$ of the HES resonance at the Fermi level. Parameters: $a=5$, $U=10\,|B|$, $\Delta=0.3\,|M|$, $\mu=0.01\,|M|$.}
\label{fig_7}
\end{figure}

In addition the characteristic temperature strongly decreases $T_s$ with increasing $d$. This is obviously explained by the decrease in the tunneling matrix. We have studied the relationship between the temperature $T_s$ and the width of resonances. First, we have found that $T_s$ can be estimated from the maximum of the temperature dependence of $\Delta G_{\Phi}(T)$ for an isolated defect with an energy level equal to the Fermi energy. An example of such a dependence is shown in Fig.~\ref{fig_2}(b). Above the maximum, the temperature dependence of $\Delta G_{\Phi}(T)$ is approximated as $1/T$. As will be shown in Sec.~\ref{Sec4}, it is precisely for this reason that a weak temperature dependence of $\Delta G$ appears in the case of scattering by many defects. $T_s$ as a function of $d$ is shown in Fig.~\ref{fig_7}(a). If we now take into account that the width of the resonances $\gamma$ also depends on the length $d$, then it is possible to exclude $d$ and obtain the dependence of $T_s$ on $\gamma$. The result of such a calculation for resonance energy equal the Fermi energy is shown in Fig.~\ref{fig_7}(b). It is clear that $T_s$ is simply proportional to the $\gamma$. Thus, the temperature $T_s$ is of the order of the width of the resonance with an energy equal to the Fermi energy.

\section{Discussion and concluding remarks}\label{Sec4}

We have studied backward scattering of electrons in helical edge states in 2D TIs with broken axial spin symmetry by isolated point defects creating bound states. 

The key point of our approach to this problem is the description of the scattering process on the basis of composite HESs formed due to the tunneling coupling between conventional HESs and bound states. The backscattering occurs as a result of two-particle scattering of electrons in these states. This approach is well suited for the case of a weakly interacting electrons in the presence of defects.

Within the frame of this approach, we have calculated the four-rank spinors of the conventional HESs and the bound states, as well as the tunneling matrix that couples the edge and bound states. It was important to find the correct dependence of the tunneling matrix on the electron momentum. Using the tunneling matrix, we have calculated the probability of e-e scattering with a `spin' flip of one of the electrons and the deviation from the quantized conductance caused by this process. In this way, we have studied the backscattering problem without using model assumptions about the tunneling coupling between the edge states and defect. Nevertheless, we have limited our study to defects with one energy level of bound states and thus excluded from consideration the electronic transitions between the energy levels of the defect, in contrast to the puddle model~\cite{PhysRevLett.110.216402,PhysRevB.90.115309}, where such transitions are fundamentally important. 

The composite HESs have two important features compared to conventional HESs. First, their local density of states near the defect has a resonance, the energy of which is somewhat shifted relative to the energy of the bound state and the width is determined by the tunneling matrix. Second, the wave functions of composite HESs are much more complicated. They contain three components with different spatial distribution. One of them is a propagating wave, which at infinity coincides with the conventional HES up to phase. The other component is localized directly near the defect and is composed of both Kramers partners of bound states of an isolated defect. The third component is a cloud around a defect formed by a wide set of wave functions composed of both Kramers partners of conventional HESs. Its amplitude decays slowly with the distance from the defect. Accordingly, there are three components of the matrix element of the e-e interaction potential, which determines the backscattering. To be specific we have studied the scattering process in which one of two colliding electrons is scattered back.

The backscattering decreases the ballistic conductance with respect to the quantum $e^2/h$. The conductance deviation $\Delta G$ from this quantum also contains three components, each of which has an own type of temperature dependence. Their relative contribution to the magnitude of the total effect is largely determined by the tunneling matrix and, in particular, by its dependence on the electron momentum. 

The backward scattering caused by the propagating-state component of the wave function is due to electron transitions between Kramers partners of the conventional HESs with different momenta, which are possible because the axial spin symmetry in broken. An important role in this process is played by the function $g_{k',k}$, which determines the non-orthogonality of the Kramers partners with different $k$ (see Eq.~(\ref{g})). In this case, the role of defects is insignificant. It is reduced only to a change in the phase that the wave function acquires when an electron passes a defect. This component of the conductance deviation, $\Delta G_{\Psi}$, rather strongly depends on the temperature, $\Delta G_{\Psi}\sim T^5$, and is small compared to the conductance deviation caused by a single defect in the channel of reasonable length.

The bound-state component plays a leading role in suppressing the conductance and largely determines its temperature dependence. In the case when electrons are scattered by only one defect, the conductance deviation due to the bound-state component $\Delta G_{\Phi}$ has a peak as a function of the Fermi level. At low temperatures $T\lesssim \gamma_{\mathcal{K}_0}$ its width is determined by the width of the resonance of composite HESs, but at higher temperatures the peak expands significantly. As a function of temperature, $\Delta G_{\Phi}$ first increases as $\sim T^4$ at low temperatures, $T \lesssim \gamma_{k_0}$, reaches a maximum with increasing temperature, and then decreases approximately as $1/T$.

The cloud component of the conductance deviation $\Delta G_{\Psi\Omega}$ also has a maximum as a function of the Fermi energy, but it is much wider than the peak of $\Delta G_{\Phi}$. Thus, in comparison with the bound-state component, the cloud component prevails far from the resonance energy and is small near it. Another feature is that $\Delta G_{\Psi \Omega}$ always increases significantly with temperature, and therefore becomes important at high temperatures.

The total deviation of the conductance $\Delta G$ created by all components, as a function of the Fermi level, has a maximum near the resonance of the composite HESs, where the bound-state component $\Delta G_{\Phi}$ predominates. However, the width of this maximum significantly exceeds the width of the resonance of the local density of composite HESs and increases with temperature. Out of the resonance, the conductance deviation is determined by the cloud component $\Delta G_{\Psi \Omega}$.

These results are generalized to the case when there are many defects, the binding energy of which is distributed over the gap. For simplicity, they are considered as uncorrelated scatterers. We have found that in the case of a uniform or slowly varying distribution of the energy levels of defects over the gap, the temperature dependence of the conductance deviation changes as follows. At low temperatures, $\Delta G$ increases rapidly in a narrow temperature range, but then $\Delta G$ changes very slowly over a wide temperature range. The characteristic temperature $T_s$, above which $\Delta G$ weakly depends on temperature, is estimated by the width of the resonance of composite HESs with resonance energy equal to the Fermi energy, $T_s\sim\gamma_{\mathcal{K}_0}$. 
In the regime of weak temperature dependence, at $T>T_s$,  the conductance suppression is mainly due to  the bound-state component of the backscattering process, and $\Delta G_{\Phi}$ dominates in the total conductance deviation $\Delta G$. In this case, the variation of $\Delta G$ with temperature depends on the width of the resonances. A decrease in the resonance width leads to a weaker temperature dependence of $\Delta G$ and a decrease in the temperature $T_s$.

Let us make several numerical estimates for conditions close to experiment. As an example, we put $d\simeq 18\sqrt{|B/M|}$ to estimate what happens when the resonance is narrow. In this case Fig.~\ref{fig_7} shows that $T_s\simeq 2\cdot 10^{-3}|M|$. If we use parameters of HgTe/CdHeTe heterostructures, this corresponds to $d\simeq$150~nm and $T_s\simeq 2\cdot 10^{-2}$~K. Thus we can expect that at $T> 2\cdot 10^{-2}$~K the conductance deviation very weakly depends on temperature as shown by the red line in Fig~\ref{fig_6}(b), for $d=17\sqrt{|B/M|}$. This estimate is close to the lower boundary of the temperature range where the temperature independent conductance was observed experimentally in HgTe/CdHeTe heterostructures~\cite{PhysRevB.89.125305}.

The physical reason for the weak temperature dependence of the conductance deviation stems from the fact that the conductance deviation created by one defect $\Delta G_1$ decreases with temperature approximately as $1/T$ at $T>\gamma_{\mathcal{K}_0}$. In this case, an increase in the temperature leads to such a decrease in $\Delta G_1$ that is exactly compensated by the increase in the number of effectively scattering defects. Indeed, effectively scattering defects are those whose energy lies in a layer with a width of the order of $T$ around the Fermi level. Thus, the product of the partial conductance deviation per defect and the layer width is nearly constant.    

More precisely this can be seen from Eq.~(\ref{delta_G_phi_simple}), which approximately describes $\Delta G_{\Phi}$ for a single defect at $T>\gamma_{\mathcal{K}_0}$. In the case of many defects with energy levels distributed with the density $\rho(\varepsilon_0)$, the correction to quantized conductance reads
\begin{equation}
\frac{\Delta G_{\Phi}}{G_0}=\int_{-vK_c}^{vK_c} \frac{d\varepsilon_0}{T} \mathcal{D}(\varepsilon_0)\rho(\varepsilon_0)\frac{2e^{3(\varepsilon_0+\Sigma-\mu)/T}}{\left[1+e^{(\varepsilon_0+\Sigma-\mu)/T}\right]^{5}}\,.
\end{equation}
It is easy to see that in the integrand, the last term containing exponential functions has a sharp peak of the width about $T$ wide, and other functions change slowly on this scale. So, the integral is simplified as follows
\begin{equation}
\frac{\Delta G_{\Phi}}{G_0}\approx 2\mathcal{D}(\mu-\Sigma)\rho(\mu-\Sigma)\int_{-\frac{|M|-\Sigma+\mu}{T}}^{\frac{|M|+\Sigma-\mu}{T}}dt\frac{e^{3t}}{\left(1+e^{t}\right)^{5}}\,.
\end{equation} 
Here the integral is almost constant and equals approximately 1/12, when the Fermi level lies not too close to the band edges, $|M|-|\mu|\ll T$, and the temperature is not very high, $T\ll |M|$. Thus, $\Delta G_{\Phi}$ varies slightly with temperature, but the specific form of its temperature dependence is determined by the distribution function $\rho(\varepsilon_0)$ of defect energy levels and the energy dependence of the tunneling matrix. If $\rho(\varepsilon_0)$ and $w_{k,\sigma;\lambda}$ vary significantly with energy, the temperature dependence of $\Delta G_{\Phi}$ is modified and can be both slowly increasing and decreasing.  

The lower boundary $T_s$ of the region of a weak temperature dependence of the conductance increases significantly with decreasing distance $d$ between the defect and the edge, and the temperature dependence of $\Delta G$ at $T>T_s$ noticeably increases. This suggests that the weak temperature dependence of the conductance observed down to low temperatures may be caused by defects located far enough from the edge. In this regard, it should be noted that the situation where a defect is located close to the edge requires the use of other theoretical approaches, since the coupling between the defect and the edge states is not weak. In this case, the HESs are strongly deformed, forming a flow around the defect, as was shown with using the nonperturbative approach~\cite{PhysRevB.91.075412}.

Now let us estimate the magnitude of the conductance deviation at temperatures above $T_s$. For the parameters of heterostructures HgTe/CdTe, it is seen from Fig.~\ref{fig_6} that $\Delta G/(NG_0)\simeq (5 \div 8)\cdot 10^{-3}$. Therefore, to halve the conductance, the number of defects in the source-drain space should be $N\simeq$80. The source-drain distance is typically about $10^{-2}$\,cm~\cite{PhysRevLett.123.047701}. Thus the average distance between the defects, which determine the scattering, is estimated as $\sim 10^{-4}$\,cm. This means that a low enough density of effectively scattering point defects is sufficient to cause a fairly strong suppression of conductance over a wide temperature range.

In conclusion, we briefly summarize the proposed mechanism of the breakdown of topological protection. The mechanism assumes the presence of a sufficiently strong SOI breaking the axial spin symmetry, point defects with at least one discrete energy level, and a weak e-e interaction. The key role is played by the fact that, due to the tunneling coupling of edge states with a defect, a Kramers pair of composite HESs is formed, in which all wave functions of both Kramers pairs of conventional HESs and bound states are mixed. The interaction of electrons in precisely these states leads to transitions between the Kramers partners of HESs even in the presence of only one energy level of bound states.

For backscattering to occur, it is important that the tunneling matrix contains two (in the case of a defect with one level) independent components, $w_{k,+;+}$ and $w_{k,+;-}$, which describe the mixing of all partners of two Kramers pairs of conventional HESs and bound states. It can be said that backscattering occurs as a result of electronic transitions mainly between composite HESs with close energies lying near the energy level of an isolated defect. It is for this reason that the magnitude of the backscattering effect is largely determined by the derivative of the tunneling matrix elements with respect to the wave vector of electrons. The absence of one of the matrix components, $w_{k,+;-}$, as well as the absence of a dependence of tunneling matrix elements on $k$, makes impossible the mechanism of backscattering we are studying. This mechanism leads to a weak temperature dependence of conductance, when scattering occurs on many defects with energy levels distributed almost uniformly over a wide energy range in the band gap.

Conceptually close backscattering mechanism was developed in Refs.\cite{PhysRevLett.110.216402,PhysRevB.90.115309} for another, much more complex system of electronic puddles, which have a discrete energy spectrum and are tunnel-coupled with edge states. Therefore, it is interesting to explain the most fundamental physical differences of this mechanism from ours, although any detailed comparison is incorrect, since both systems under study are too different. A puddle is considered as a quantum dot with large enough number of energy levels. The key point that determines the backscattering is electronic transitions between the energy levels of the quantum dot with changing the Kramers index of bound states, which occur as a result of e-e interaction in the quantum dot when axial spin symmetry is broken. In our case, the point defect has only one energy level. Therefore, there are no electron transitions in such a dot as the point defect and the mechanism proposed for puddles does not work. On the contrary, the mechanism proposed here can work in the case of puddles.

There is another aspect. In the puddle model, the tunneling transitions between the edge and bound states are considered under the assumption that the Kramers index, R or L, is conserved. This is not possible for a point defect with one or even several levels, although it might be justified for a large quantum dot. Accordingly, the tunneling coupling is described by only one matrix element for each level, which, moreover, is independent of the wave vector. In this way, the backscattering mechanism we have studied is lost. The conductance suppression produced by this mechanism, as we have shown, is sufficiently large, at least for point defects, although the situation may change in the case of large puddles.  

Regarding the problems raised in experiments, the analysis presented here allows one to obtain a satisfactory quantitative estimate of the conductance deviation, starting directly from the Hamiltonian with known material parameters and without using phenomenological parameters. Another significant problem is the weak temperature dependence of the conductance suppression. Our analysis shows that this effect can be explained by backscattering from many point defects with energy levels scattered over a wide range. This mechanism can be a hypothesis for the interpretation of experiments, but has no direct experimental confirmation, like other mechanisms based on inelastic e-e scattering. An unsolved problem is the behavior of $\Delta G$ in the low temperature limit. The available experiments do not reveal a sharp drop in $\Delta G$ at $T\to 0$, as is the case in our theory and in other theories based on inelastic e-e scattering~\cite{PhysRevB.90.115309}. Obviously, other models are required to solve this problem, such as models with spontaneous spin symmetry breaking~\cite{PhysRevLett.122.016601}.

\begin{acknowledgments}
This work was carried out in the framework of the state task and supported by the Russian Foundation for Basic Research, project No.~20-02-00126.

\end{acknowledgments}

\appendix
\section{Tunneling matrix}\label{App1}

The tunneling matrix $w_{k,\sigma;\lambda}$ plays key role in the backscattering mechanism studied in this paper. Of essentially importance is its dependence on the electron momentum in the HESs. Therefore, the tunneling matrix should be carefully calculated.

The tunneling Hamiltonian coupling edge and bound states was derived in Ref.\cite{PhysRevB.102.075434} where it was found that $H_T=-H_{\rm{bulk}}$, with $H_{\rm{bulk}}$ being the bulk Hamiltonian of 2D material. The matrix elements $w_{k,\sigma;\lambda}$ have the form
\begin{equation}
 w_{k,\sigma;\lambda}=-\varepsilon_0\langle k,\sigma|\lambda\rangle - \langle k,\sigma|V|\lambda\rangle\,,
\label{tunnel_matrix}
\end{equation} 
where $V=V(x,y-d)$ is the potential of the defect.

As the bulk Hamiltonian we have used the Hamiltonian of the BHZ model~\cite{Bernevig1757} generalized by including the SOI due to bulk inversion asymmetry~\cite{doi:10.1143/JPSJ.77.031007}. The BHZ Hamiltonian in the commonly used notations reads
\begin{equation}
H_{\rm{BHZ}}(\mathbf{\hat{k}})=\begin{pmatrix}
 M\!-\!B \hat{k}^2 & A\hat{k}_+ & 0 & -\Delta \\
 A\hat{k}_- & -M\!+\!B \hat{k}^2 & \Delta & 0\\
 0 & \Delta & M\!-\!B \hat{k}^2 & -A\hat{k}_-\\
 -\Delta & 0 & -A\hat{k}_+ & -M\!+\!B \hat{k}^2 
\end{pmatrix}\,,
\label{HamiltonianBHZ}
\end{equation}
where $\mathbf{\hat{k}}$ is the momentum operator, $\hat{k}_{\pm}=\hat{k}_x\pm i\hat{k}_y$. The Hamiltonian is presented in the basis of electron-like and heavy-hole states with spin up and down $\left(|e\uparrow\rangle,|h\uparrow\rangle,|e\downarrow\rangle,|h\downarrow\rangle\right)^T$. The parameters $M$, $A$, $B$ are well known for specific materials. $A$ describes the hybridization of the electron and hole  basis states, $M$ is the mass term, and $B$ is the parameter of the dispersion in the electron and hole bands, which are assumed to be symmetric.  SOI is presented by the parameter $\Delta$. An important role belongs to the parameter $a=A/\sqrt{|BM|}$. 

The wave functions are spinors of the fourth rank. This somewhat complicates the calculations, but it is impossible to lower the rank of spinors, because in a system with broken axial symmetry, the spin projection onto any axis is not a quantum number.

Specific numerical calculations of the tunneling matrix and the conductance deviation from the quantized value, are carried out using the BHZ model parameters close to HgTe/CdTe heterostructures: $M$=-0.01~eV, $B$=-0.7~eV\,nm$^2$, $A$=0.37~eV\,nm. The SOI parameter is taken equal to $\Delta=0.3|M|$, which is a reasonable theoretical estimate~\cite{WINKLER20122096,PhysRevB.93.075434}. For HgTe/CdTe heterostructures, the parameter $a$ is estimated as $a$=4.4. In our numerical calculations we use two values of $a$: $a$=5, which is close to HgTe/CdTe, and $a$=2, which allows us to understand how the hybridization of the electron and hole bands affects the backscattering.

The wave functions of the conventional HESs are determined from the Schrödinger equation $H_{BHZ}\Psi_{k,\sigma}=\varepsilon_{k,\sigma}\Psi_{k,\sigma}$ with the boundary condition $\Psi_{k,\sigma}(x, y=0)=0$. They are found using the technique described earlier~\cite{PhysRevB.102.075434}.

The wave functions of the bound states are found from the Schr\"odinger equation $[H_{\rm{BHZ}}+V(r)]\Phi_{\lambda}=\varepsilon_{0}\Phi_{\lambda}$. The eigenenergy $\varepsilon_0$ and the spinors $\Phi_{\lambda}$ are calculated in the case of short-range potential $V(\mathbf{r})=v_0\Lambda^2/\pi \exp(-\Lambda^2r^2)$ with amplitude $v_0$. The radius $\Lambda^{-1}$ of the potential is supposed to be small. The calculation method is described in Refs.~\cite{doi:10.1002/pssr.201409284,SABLIKOV20161}. 

\begin{figure}
\includegraphics[width=0.8\linewidth]{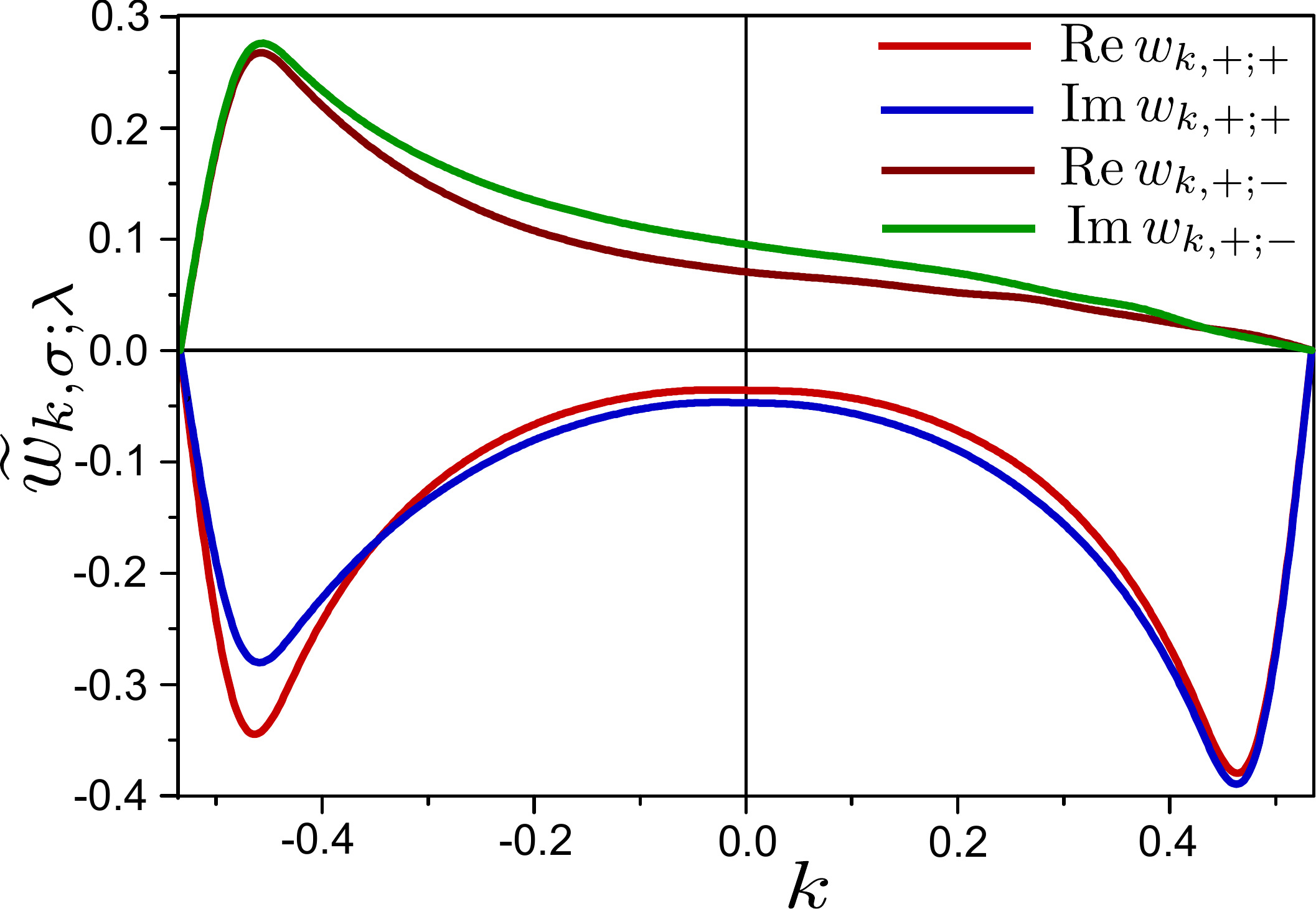}
 \caption{Tunneling matrix elements $\widetilde{w}_{k,+;+}$ and $\widetilde{w}_{k,+;-}$ as functions of the wave vector. The matrix elements are normalized to $|M|$. The wave vector $k$ is normalized to $\sqrt{M/B}$. Numerical parameters used in the calculations are $\Delta=0.3 |M|$, $a=2$, $d = 6\sqrt{B/M}$. The defect potential is chosen so that $\varepsilon_0=0.03\,|M|$.} 
\label{fig_8}
\end{figure}

The tunneling matrix elements found in this way is shown in Fig.~\ref{fig_8} for $a=2$. In the case of $a=5$, the results are qualitatively similar.

~
\bibliography{back_scatter}

\end{document}